\documentclass[11pt]{article}
\usepackage[english]{babel}
\usepackage[a4paper,margin=1in]{geometry}
\usepackage[utf8]{inputenc}
\usepackage{multirow} 
\usepackage{graphicx,float,lipsum}
\usepackage{xcolor}
\usepackage{amsmath, amssymb}
\usepackage{bbold}
\usepackage[colorlinks=true, allcolors=blue]{hyperref}
\usepackage[font=small, labelfont=bf, margin=5mm]{caption}

\newcommand{\Ll}{\mathcal{L}}

\usepackage{amsthm}
\usepackage{mathtools}          
\usepackage{enumitem}           
\usepackage{listings}           
\usepackage{todonotes}          
\usepackage{hyperref}           
\usepackage{dirtytalk}
\usepackage{amssymb}

\title{\textbf{The massive BMS character in 3D quantum gravity}}
\author{\textbf{T. Mursheed Amith}\footnote{Email: \href{mailto:tuanmursheed123@gmail.com}{tuanmursheed123@gmail.com}} \qquad \textbf{Alicia Castro}$^{a,}$$^b$\footnote{Email: \href{mailto:blanca-alicia.castro-bermudez@u-bordeaux.fr}{blanca-alicia.castro-bermudez@u-bordeaux.fr}}\\[3mm]
{\small $^a$  LIGM, CNRS UMR 8049, Université Gustave Eiffel, Champs-sur-Marne, France}\\
{\small $^b$ LaBRI, Université de Bordeaux, Talence, France.}}
\date{ }
\usepackage{graphicx}

\begin{document}

\maketitle
\begin{abstract}
We derive the one-loop partition function for three-dimensional quantum gravity in a finite-radius thermal twisted flat space with a conical defect, reproducing the massive BMS$_3$
character. We perform the computation in both discrete and continuum geometry formulations, showing consistency between them. In the discrete case, we integrate out bulk degrees of freedom in a Regge gravity framework, while in the continuum, we construct a dual non-local boundary field theory encoding geodesic length fluctuations. Our study shows that the additional modes of the massive character, compared to the vacuum case, originate from the explicit breaking of radial diffeomorphism symmetry by the defect. This provides a concrete geometric mechanism in Regge gravity, tracing the appearance of massive BMS$_3$ particles to diffeomorphism breaking by conical defects, and highlights the broader relevance of discrete geometry approaches to quantum gravity with matter.
\end{abstract}

\tableofcontents

\section{Introduction} 

What describes a particle in quantum gravity (QG)?
In quantum field theory (QFT), a particle is characterized by the irreducible unitary representations of the Poincaré group, leading to particles being labeled by spin and mass. However, in quantum gravity, where the structure of spacetime highly fluctuates, fundamental questions such as: How do we label a particle, and what do mass and spin signify in such scenarios, arise.

A partial answer emerges when, even though spacetime can fluctuate, its asymptotic structure remains fixed. For instance, for $D$-dimensional asymptotically anti-de Sitter (aAdS) spacetimes, the asymptotic symmetry group is $SO(2,D-1)$. For $D=3$, this group is composed of two copies of the Virasoro group \cite{Brown:1986nw}, and its irreducible unitary representations are labeled by their highest weight and central charges \cite{Goddard:1986ee}. These serve to build the invariant labels of mass and spin. In this framework, the notion of surface charges from General Relativity (GR) \cite{Barnich:2001jy} corresponds to the Noether charges associated with conformal transformations labeled by the Virasoro central charges. This forms one of the foundations of the AdS/CFT correspondence, where a $D$-dimensional aAdS gravitational system can be studied via its holographic conformal field theory (CFT) living on its $(D-1)$-dimensional asymptotic boundary. 

For asymptotically flat spacetimes, the symmetry group is an extension of the Poincaré group, known as the BMS group \cite{Bondi:1962px, Sachs:1962BMSORIGINAL}. This symmetry group consists of the transformations that preserve the asymptotic structure of the metric at null infinity and includes supertranslations and superrotations. In particular, in three dimensions, BMS$_3$ can be expressed as   
\begin{equation*}
    \mathrm{BMS}_3\equiv \mathrm{Diff}^+(S^1)\ltimes \mathrm{Vect}(S^1),
\end{equation*}
where $\mathrm{Diff}^+(S^1)$ denotes the group of orientation-preserving diffeomorphisms of the circle (superrotations), while $\mathrm{Vect}(S^1)$ is the abelian additive group of vector fields on the circle (supertranslations).

Despite its long history, the irreducible unitary representations of the BMS group in three dimensions were constructed only a decade ago \cite{Oblak:2016Oblackthesis}. In particular, the characters of these unitary representations are given by
\begin{equation}
    \chi_{0,j}=e^{i\alpha^0(-c/24)}\prod_{n=2}^\infty\frac{1}{|1-q^n|^2},\label{bms3massless}
\end{equation}
for the so-called vacuum character, while the massive character is given by
\begin{equation}
    \chi_{m,j}=e^{i\alpha^0(m-c/24)}\prod_{n=1}^\infty\frac{1}{|1-q^n|^2},\label{bms3massive}
\end{equation}
where $q=\exp{(i\theta)}$, $m$ is the BMS mass, $c$ is the central charge and $\alpha^0$ denotes the zeroth Fourier mode of a supertranslation. As asymptotically flat spacetimes are thought to provide a more realistic model of our universe, these results offer a promising framework for addressing the foundational questions in quantum gravity posed at the beginning of this paper. In particular, they have inspired efforts to adapt holography techniques, originally developed for AdS spacetimes, to the asymptotically flat case.

Holography plays a key role in three-dimensional quantum gravity. Since 3D pure gravity has no local degrees of freedom, physical information can only be stored in a quasi-local or global manner. This is a defining feature of Topological Field Theories (TFTs) \cite{Witten:19883DGRAVITY, Carlip:1998BOOK}. In TFTs, bulk physics can be encoded almost trivially in a boundary action. This follows from the structure of the Hamilton-Jacobi function, which receives contributions only from boundary terms \cite{Oeckl:2003vu}. Notably, in this setting, the bulk need not have a fixed geometry (e.g., AdS), and the boundary does not have to be asymptotic.

In fact, finite-distance boundaries have proven to be essential for capturing the full information of 3D quantum gravity, including contributions that vanish asymptotically \cite{Dittrich:2018xuk,Goeller:2019zpz,Freidel:2021ajp,Freidel:2023bnj}. In particular, in \cite{Freidel:2021ajp,Freidel:2023bnj}, finite-boundaries were used in the construction of a quantum symmetry algebra for 3D quantum gravity. Along these finite-distance boundaries, the presence of corners motivated a discretized/simplicial geometry picture of spacetime.

In the quantum gravity landscape, discrete/simplicial geometry approaches, where spacetime is built from elementary building blocks, are particularly useful because the presence of a coordinate-independent UV cutoff makes them intrinsically non-perturbative. Furthermore, the discreteness of spacetime provides both a numerical and combinatorial framework for computing the quantum gravity path integral \cite{ambjorn2009geometry}. For instance, numerical studies in 3D Causal Dynamical Triangulations have found evidence of a ground state for the gravitational Hamiltonian \cite{Ambjorn:2000dja}. In Loop Quantum Gravity, the Ponzano-Regge spin foam model \cite{Barrett:2008wh} provides a state sum formulation of the gravity path integral over discretized geometries. More recently, discrete geometry techniques have also been applied to AdS$_3$/CFT$_2$ \cite{Jafferis:2024jkb}, where the discrete building blocks emerge from the boundary theory data of Liouville CFT. A central framework in these approaches is Regge gravity \cite{Regge:1961px}, which discretizes general relativity using simplicial geometry. In three dimensions, the topological nature of gravity ensures discretization invariance, meaning that the partition function remains unchanged under refinements of the discretization (coarse-graining) \cite{dittrich2012path}.

The role of finite-distance boundaries in 3D Regge quantum gravity was initially explored in \cite{Bianca:2015}, where the partition function was expressed in terms of quantum length fluctuations in a solid torus geometry. The one-loop partition function was computed, resulting in the \textit{vacuum} BMS$_3$ character in the continuum limit. This work also computed the Hamilton-Jacobi function, revealing a dual boundary field theory with a Liouville-type coupling to the boundary metric. The results were later confirmed in the continuum setting \cite{asante2019holographic}, where metric fluctuations were expressed as non-local geodesic lengths from the center of the solid torus to the finite-distance boundary. Furthermore, the radial geodesic length was identified as the scalar field in the boundary field theory. These studies extended previous results on the 3D QG one-loop partition function with asymptotic boundaries, both in flat spacetime \cite{Barnich:2015Partitionfunctionatoneloop} and AdS$_3$ \cite{Maloney:2007ud,Giombi:2008vd} in vacuum. 

However, in AdS$_3$/CFT$_2$, introducing point particles has proven essential, as they represent additional saddle points of the partition function. These saddles correspond to BTZ black holes \cite{Banados:1992wn} and they are realized as conical defect geometries \cite{Krasnov:2000PointParticleinADS,Maxfield:2020ale}. Conical defects in AdS$_3$ were later incorporated \cite{Benjamin:2020mfz} leading to a one-loop partition function that reproduces the \textit{massive} Virasoro character.

Building on these developments, a natural question arises: can a simple model in flat spacetime, computable in both discrete and continuous geometry settings, reproduce the \textit{massive} BMS$_3$ character and extend previous results beyond the vacuum case? To explore this, we further develop \cite{Bianca:2015, asante2019holographic} to flat spacetimes with a conical defect representing a massive point particle. 

\begin{figure}[h]
  \centering
    \includegraphics[width=\textwidth]{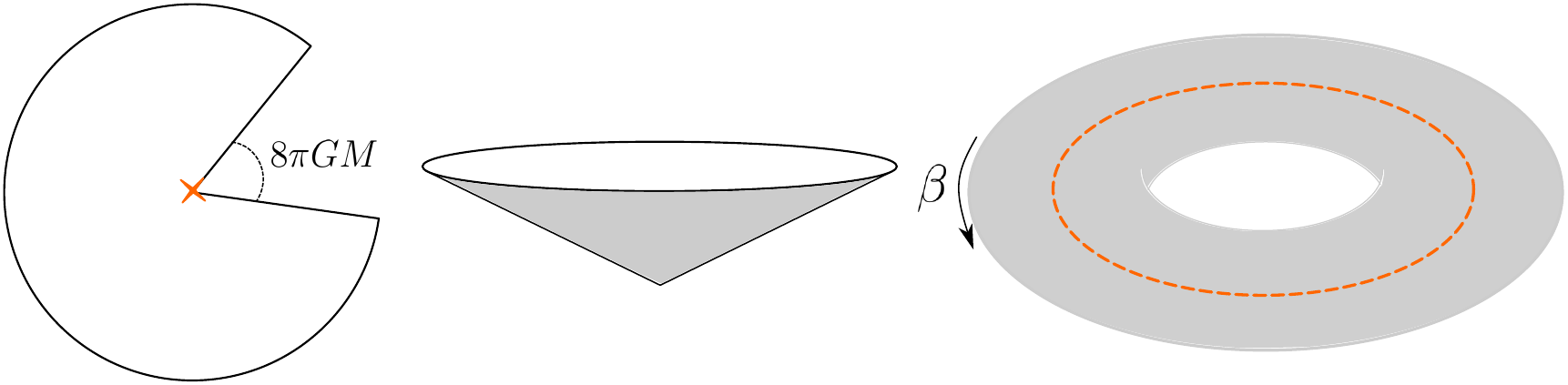}
     \put(-300,0){a)}
     \put(-90,0){b)}
  \caption{a) Cone geometry with deficit angle $8\pi GM$. b) Twisted thermal flat space with inverse temperature $\beta$. The point particle's worldline is shown in orange.}\label{cone}
\end{figure}

The classical action of a massive point particle coupled to 3D gravity is given by
\begin{equation}
    S_p=-\frac{1}{16\pi G}\left(\int_{\mathcal M} d^3x\sqrt{g}R+2\int_{\partial \mathcal{M}} d^2y\sqrt{h}K\right)-M\int_{\mathcal{WL}} d\tau \sqrt{\dot{x}^\mu\dot{x}^\nu g_{\mu\nu}(x)}.\label{actionpointpart}
\end{equation}
The first term corresponds to Einstein-Hilbert's action for $\Lambda=0$ and the second term corresponds to the Gibbons-Hawking-York boundary term. The third term corresponds to the worldline ($\mathcal{{WL}}$) of a point particle with rest mass $M$ in a gravitational background.
The equations of motion of \eqref{actionpointpart} are
\begin{equation}
\label{eqnofmotion1}
    R_{\mu\nu}-\frac{1}{2}Rg_{\mu\nu}=8\pi GT_{\mu\nu},
\end{equation}
where $T_{\mu\nu}$ is the energy-momentum tensor of a massive point particle in its rest frame. In cylindrical coordinates, this is
\begin{equation}
    T_{00}=M\delta^{(2)}(\bar{r}),\hspace{3mm}T_{i0}=0=T_{ij},
\end{equation} 
with $i\in\{1,2\}$.
Thus, the solution for the metric reads
\begin{equation}
\label{metricnocone}
    ds^2=(1-4GM)^2dt^2+(1-4GM)^2dr^2+r^2d\theta^2.
\end{equation}
Rescaling the radial and time directions by $(1-4GM)$, this results in the following spacetime metric \cite{Deser:19833Dpointparticlessolution}
\begin{equation}
ds^2=dt^2+dr^2+r^2d\theta^2,\label{metriccone}
\end{equation}
with $0 \leqslant \theta\leqslant 2\pi(1-4GM)$ and $0 \leqslant t \leqslant \beta$. For any fixed time, this metric describes a cone with deficit angle $8\pi GM$ (see Fig. \ref{cone}). 
To account for the symmetries of this orbifold, one usually considers a more general  periodic identification $(r,t,\theta) \sim (r, t + \beta , \theta + \gamma )$ for $\gamma>0$. This geometry is called Twisted thermal flat space with inverse temperature $\beta$.
\\
Thus, a point particle coupled to 3D gravity deforms the vacuum flat solution solely by introducing a conical defect. This provides a non-trivial test case that incorporates massive degrees of freedom while preserving the topological nature of 3D gravity. Crucially, this setup admits a discrete formulation since the partition function remains discretization-invariant, allowing us to study the model using Regge calculus. Although the role of point particles in the quantum path integral of 3D gravity remains less clear in flat spacetime, it is structurally analogous to conical defects and BTZ black holes in AdS$_3$. 

The central result of this work is that, by extending the framework of \cite{Bianca:2015, asante2019holographic} to include a point particle, we reproduce the massive BMS$_3$ character and identify its microscopic origin in discrete geometry. In particular, we show that the additional contributions in the massive character arise precisely from the breaking of diffeomorphism invariance in the Regge discretization. This provides a concrete and computable example of how physical degrees of freedom can emerge from gauge symmetry breaking in Regge gravity. Remarkably, we confirm this mechanism independently in the continuum, offering a robust cross-check. Our results thus establish a clear benchmark for tracking diffeomorphism breaking in 3D Regge gravity and highlight the broader utility of discrete methods in quantum gravity with matter.

This paper begins by constructing the Regge action for a point particle in flat space in Section \ref{section: Discrete}. Later, we compute the partition function of 3D gravity with a point particle in the \textit{discrete} Regge gravity framework, demonstrating how the massive BMS$_3$ character emerges and highlighting the role of diffeomorphism symmetry breaking compared to the vacuum case. In Section \ref{section: comp_w_vacuum}, we compare these results with the vacuum case. Section \ref{section: Continuum} presents a direct computation of the partition function in \textit{continuum} field theory. In particular, we explain how the continuum geometry perspective highlights the non-local nature of the boundary field theory, where the geodesic length plays the role of the boundary field. In Section \ref{section: comp_w_discrete}, we compare the boundary field theories obtained in both the discrete and continuum approaches. Finally, in Section \ref{section: Discussion}, we relate our results to current research and outline future directions.

\section{Partition function in the discrete}\label{section: Discrete}
Our first goal is to compute the one-loop partition function for Regge gravity coupled to a massive point particle and verify its consistency with the massive BMS$_3$ character. In this section, we outline the steps taken to achieve this, following the procedure described in \cite{Bianca:2015}, which focused on the vacuum case. Notably, the presence of the particle does not affect the topological nature of the theory, allowing us to apply the same approach. \par

\subsection{Regge action for a point particle}
\begin{figure}[h]
  \centering
    \includegraphics[width=0.7\textwidth]{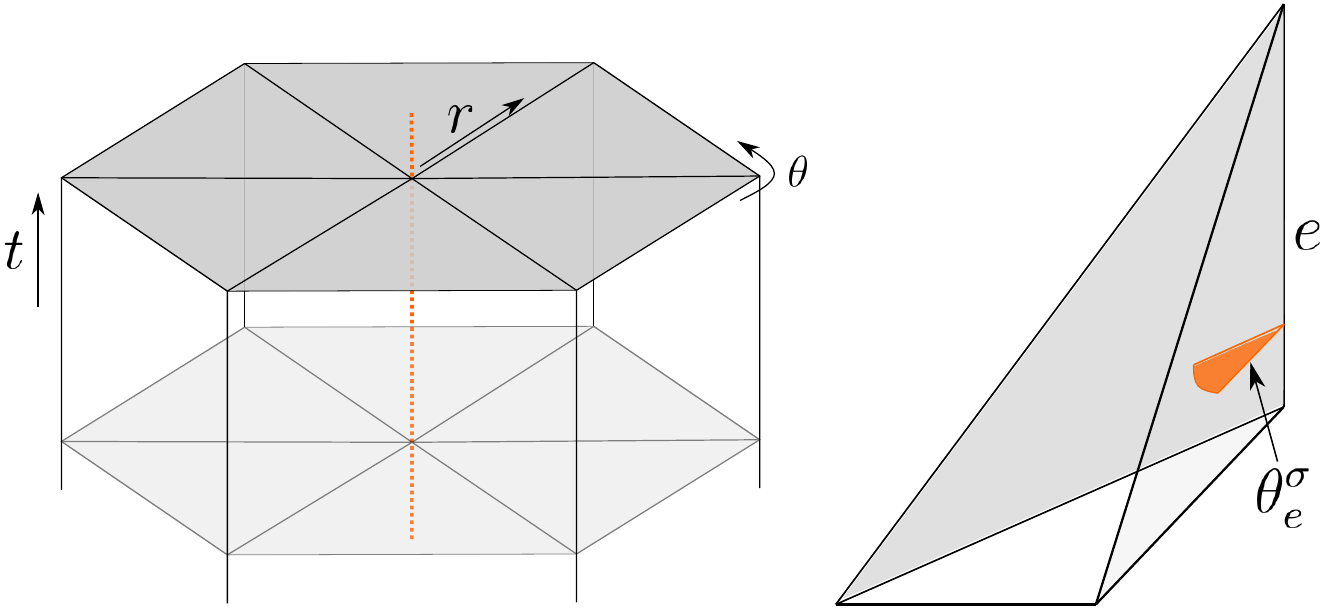}
     \put(-200,-10){a)}
     \put(-80,-10){b)}
  \caption{a) Spatial slices of the discretized torus. The orange dotted line represents the worldline of the point particle. b) A single tetrahedron $\sigma$ and the interior dihedral angle localized on the edge $e$.}
    \label{triangulation}
\end{figure}

The Einstein-Hilbert action can be reformulated in terms of simplicial geometry, which represents spacetime as a piecewise flat discrete manifold, leading to the well-known Regge action \cite{Regge:1961px} with boundaries \cite{hartle1981boundary}. In three dimensions, the Regge action takes the following form
\begin{equation}
    S_R=-\frac{1}{8\pi G}\sum_{e\subset\mathcal{T}} \epsilon_e(l_{e'}) l_e -\frac{1}{8\pi G}\sum_{e\subset\partial\mathcal{T}} \omega_e(l_{e'})l_e\label{reggeaction}
\end{equation}
where $l_e$ represents the lengths of the edges $e$ in a simplicial complex. In this work, we use triangulations $\mathcal{T}$ composed of tetrahedra. This choice is made for computational convenience, as the final result should be independent of the specific triangulation due to the topological nature of the theory. The quantities $\epsilon_e$ and $\omega_e$ denote the deficit angles in the bulk and on the boundary, respectively (see Figs. \ref{spatialslice} and \ref{triangulation}). They are expressed as follows
\begin{equation}
    \epsilon_e(l_{e'})=2\pi-\sum_{e\subset\sigma}\theta_e^\sigma(l_{e'}),\label{deficitbulk}
\end{equation}
\begin{equation}
    \omega_e(l_{e'})=\pi-\sum_{e\subset\sigma}\theta_e^\sigma(l_{e'}),\label{deficitbound}
\end{equation}
where $\theta^\sigma_e$ denotes the interior dihedral angle along the egde $e$ in the tetrahedron $\sigma$. The deficit angles encode bulk and extrinsic curvatures, respectively.

Building on the classical solution for a point particle in a 3D geometry \eqref{metriccone}, we derive its corresponding Regge action. We consider a massive particle at rest, located at the center of a torus (see Fig. \ref{triangulation} a)), and represent its discretized worldline as the sum of the lengths of the edges along its trajectory. This reads
\begin{equation}
    S_p=-\frac{1}{8\pi G}\left(\sum_{e\subset\mathcal{T^o}} \epsilon_e(l_{e'}) l_e +\sum_{e\subset\partial\mathcal{T}} \omega_e(l_{e'})l_e+\sum_{e\subset \mathcal{WL}} 8\pi GM \hspace{1mm}l_e\right),\label{reggepointpart}
\end{equation}
where the last term corresponds to the discretized worldline of a massive point particle in its rest frame. The equations of motion are obtained by varying the action \eqref{reggepointpart} with respect to the lengths $l_e$ and imposing $\delta_{l_e} S_p=0$. These are  
\begin{equation}
\left\{
\begin{aligned}
\epsilon_e & = 0 ,  & e \nsubseteq \mathcal{WL}\\
\epsilon_e & = -8\pi GM , & e\subset \mathcal{WL}
\end{aligned}
\right.
\label{deficitangle}
\end{equation}
They correspond to a discrete geometry with a conical deficit angle $8\pi GM$ along the worldline at $r=0$ and flat everywhere else. This is consistent with the continuum result \eqref{metriccone}.


\subsection{Linearized theory}
To compute the partition function of this system, we must account for quantum fluctuations of the metric. In Regge gravity, these correspond to fluctuations in the lengths of the triangulation $\mathcal{T}$ \cite{Rocek:1982tj}. For this, we use linearized Regge calculus, derived by expanding \eqref{reggepointpart} to quadratic order in length perturbations around the solution to the equation of motion \eqref{deficitangle}. Specifically, we write $l_e = L_e + \lambda_e$, where $L_e$ represents the lengths of the triangulation that solves \eqref{deficitangle}, and $\lambda_e$ denotes the length fluctuations. The linearized Regge action, coupled to a point particle and expanded up to second order, is given by
\begin{equation}
    S_p^{LR}=S_p^{(0)}+S_p^{(1)} +S_p^{(2)}.\label{slinearized}
\end{equation}
\par
The zeroth and first-order terms are given by the Hamilton-Jacobi function. These terms vanish in the bulk of the discretized torus because they are proportional to the equations of motion, but not at the boundary.
The second-order term is given by the Hessian of the action \cite{Dittrich:2007HessianFirst},
\begin{equation}
    S_p^{(2)}=\frac{1}{2}\sum_{e,e'}\frac{\partial^2S_p}{\partial l_e\partial l_{e'}}\lambda_e\lambda_{e'}=\frac{1}{16\pi G}\sum_{e,e'}\lambda_e H_{ee'}\lambda_{e'}.\label{generalhessian}
\end{equation}
The Hessian $H_{ee'}$ will be a key object in this work since it is the only term in \eqref{slinearized} that depends on bulk degrees of freedom. Since the massive part of the action \eqref{reggepointpart} is linear in length variables, the Hessian does not change its form compared to the vacuum case. This will allow us to use some of the results of \cite{Bianca:2015}.


\subsection{Symmetry and diffeomorphisms}
When considering the measure of the partition function in terms of discrete geometry variables, the challenge of translating diffeomorphism invariance from the continuum to the discrete setting arises. This difficulty, tied to the measure of integration over geometries, persists in both frameworks. In particular, in 3D Regge gravity, diffeomorphism invariance is closely related to the triangulation independence of the Regge action.\par
Linearized Regge calculus provides a significant advantage in this context, offering a well-established discrete analogue of diffeomorphism invariance \cite{Bahr:2009Reggeactionlinearization}. Moreover, it enables the construction of an invariant measure, as demonstrated in \cite{dittrich2012path}. This is
\begin{equation}
\mathcal{D}\mu(l)=\prod_\sigma\frac{1}{\sqrt{12\pi V_\sigma}}\prod_{e\in bulk}\frac{l_edl_e}{\sqrt{8\pi G\hbar}}\prod_{e\in bdry}\sqrt{\frac{l_e}{\sqrt{8\pi G\hbar}}},\label{generalmeasure}  
\end{equation}
where $V_\sigma$ is the volume of a tetrahedron. As expected, this measure is invariant under refinement refinements of the triangulation. 

This gauge symmetry shows on the null vectors of the Hessian \eqref{generalhessian}. The number of null vectors is associated with the independent directions in which the bulk vertices can be moved without changing the boundary triangulation and the volume of the Regge action. As found in \cite{Bianca:2015}, given $n$ null vectors, we want to remove the  factor
\begin{equation}
     \frac{1}{2\pi}\prod_{a=1,2,3}\frac{1}{\sqrt{8\pi G\hbar}}dx^a_p,\label{singulargaugemeasure}
\end{equation}
per bulk vertex. Here $x^a_p$ are the Euclidean coordinates of the vertex $p$ in $\mathbb{R}^3$. This measure represents the measure over the gauge orbits.\par
Even though we have seen that the form of the Hessian remains unchanged with respect to the vacuum case, its number of eigenvectors will change. This is a signal of broken diffeomorphisms.


\subsection{Triangulation}
\begin{figure}[h]
  \centering
    \includegraphics[width=0.8\textwidth]{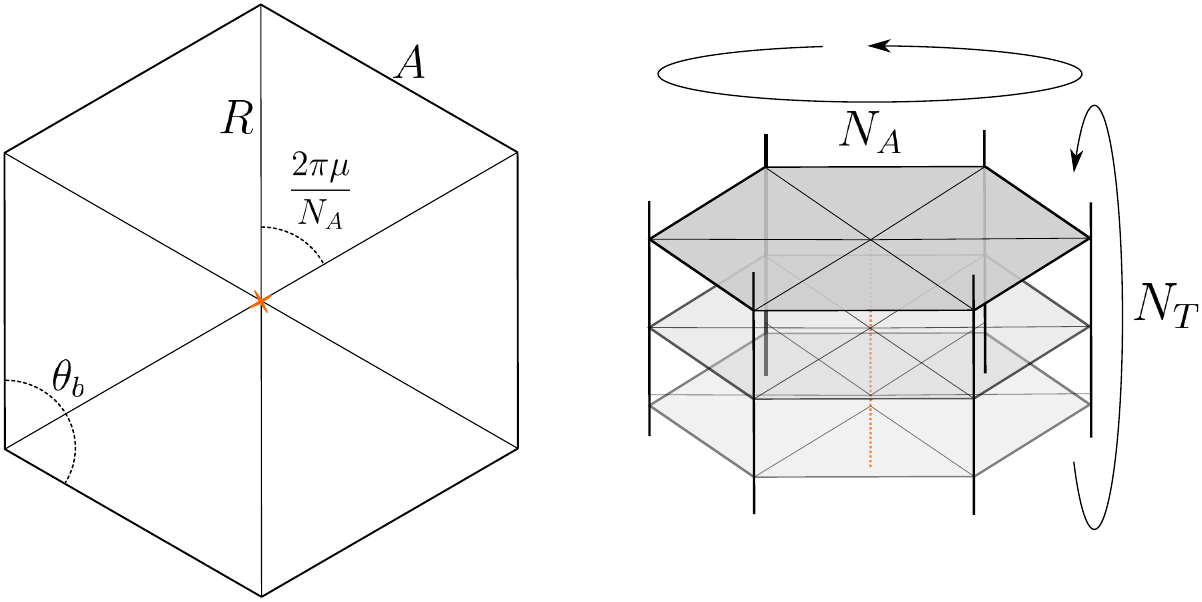}
         \put(-270,-10){a)}
     \put(-100,-10){b)}
  \caption{a) Spatial slice of the triangulation. b) Background triangulation for the torus with $N_A$ prisms in the angular direction and $N_T$ slices in the time direction.}
    \label{spatialslice}
\end{figure}

\begin{figure}[h]
  \centering
    \includegraphics[width=0.9\textwidth]{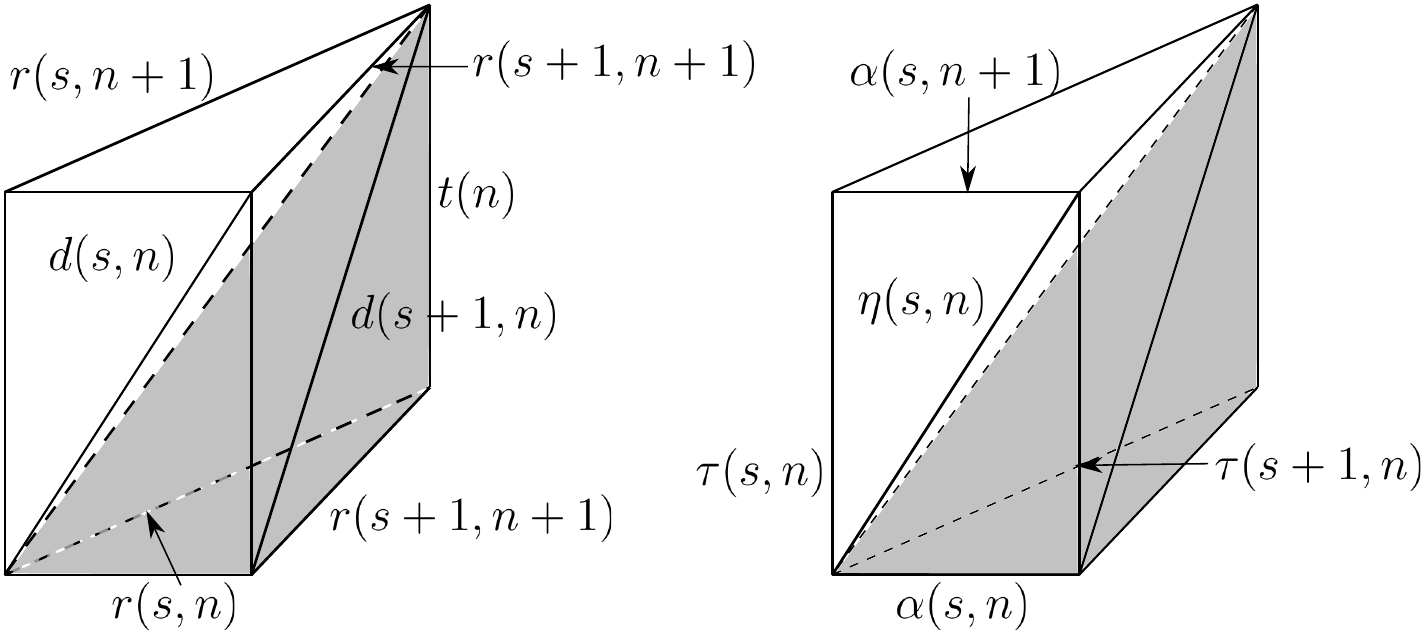}
  \caption{Subdivision of each prism in three tetrahedrons and the bulk (left) and boundary (right) fluctuations of its lengths \cite{Bianca:2015}.}  \label{discretization}
\end{figure}

To begin, we select a background triangulation for the torus. This triangulation represents a configuration with edge lengths that satisfy the equations of motion \eqref{deficitangle}. For simplicity, we use a modified version of the triangulation introduced in \cite{Bianca:2015}. The construction consists of horizontal slices along the time direction, corresponding to the non-contractible cycle of the torus. Each slice is divided into six building blocks, as depicted in Fig. \ref{spatialslice}, and each building block is further subdivided into three tetrahedra, as shown in Fig. \ref{discretization}.
The background triangulation, characterized by edge lengths $A$, $R$ and $T$, and the corresponding spatial slices are illustrated in Fig. \ref{spatialslice}.\par
Note that the deficit angle appears along the central axis. It will be useful to express the trigonometric relation between $A$ and $R$ in terms of the deficit angle. For this purpose, we define the quantity
\begin{equation}
    x:=\frac{A^2}{2R^2}=1-\mathrm{cos}\left(\frac{2\pi\mu}{N_A}\right)\label{definitionx}
\end{equation}
where $\mu=(1-4GM)$. This variable is a modified version of the one defined in \cite{Bianca:2015} for the vacuum case, and it matches that value when $\mu = 1$.
Additionally, imposing the flatness condition on each triangle in a spatial slice reflects the deficit angle in the boundary deficit angle $\theta_b$ (see Fig. \ref{spatialslice}), which is given by
\begin{equation}
    \theta_b=\pi-\frac{2\pi\mu}{N_A}.\label{thetab}
\end{equation}
After establishing the background triangulation, we now turn to studying the linearized quantum theory. To this end, we introduce length fluctuations of the edges of the triangulation, denoted by
\begin{equation}
    \lambda(s,n)=(t(n),r(s,n),d(s,n),\tau(s,n),\alpha(s,n),\eta(s,n)).\label{eq:length_fluc_s_n}
\end{equation}
The pair $(s,n)$ denotes coordinates in this triangulation corresponding to angular and temporal directions as shown in Fig. \ref{discretization}.
In these variables, we can write the partition function of the linearized action \eqref{slinearized} in the following way
\begin{equation}
    Z(\{\tau(s,n),\alpha(s,n),\eta(s,n)\})=\int\prod_ndt(n)\prod_{s,n}dd(s,n)dr(s,n)\mu(l)\mathrm{exp}\left(-\frac{S^{LR}_p}{\hbar}\right),
\end{equation}
where $\mu(l)$ is the density of the measure \eqref{generalmeasure}. This density evaluated in this background triangulation takes the form
\begin{equation}
    \mu(l)=\frac{R^{N_AN_T}T^{N_T}(R^2+T^2)^{N_AN_T}}{(8\pi G\hbar)^{N_T(2N_A+1)/2}}\frac{1}{(12\pi V_\sigma)^{3N_AN_T/2}}
    \sqrt{\frac{A_{N_AN_T}T^{N_AN_T}(A^2+T^2)^{N_AN_T/2}}{(8\pi G\hbar)^{3N_AN_T/2}}},\label{triangulationmeasure} 
\end{equation}
where background tetrahedron volume is given by 
\begin{equation}
    V_\sigma=\frac{1}{12}ATR\sqrt{4-2x}.\label{eq:vol_tetrahedron}
\end{equation}
Using the linearization of the action \eqref{slinearized}, the partition function can be written as
\begin{equation}
     Z(\{\tau(s,n),\alpha(s,n),\eta(s,n)\})=e^{-\frac{1}{\hbar}S_p^{(0)}}D\hspace{1mm}e^{-F(\{\tau(s,n),\alpha(s,n),\eta(s,n)\})},\label{formofpartfunction}
\end{equation}
where $S_p^{(0)}$ is the action \eqref{reggepointpart} on-shell, $D$ is proportional to the one-loop determinant and $F$ is the part of the Hamilton-Jacobi function linear and quadratic in boundary length fluctuations. It is worth mentioning that we expect this form for the partition function since in the vacuum case it has this structure, both in the continuum and in the discrete. In this work, we show that \eqref{formofpartfunction} gives the correct massive BMS$_3$ character both in the discrete and continuum formulations.

\subsection{Hamilton-Jacobi function}
Having established the structure of the partition function in terms of length fluctuations in the previous section, we now focus on computing the first two terms in the linearized Regge action \eqref{slinearized}. These terms are derived from the Hamilton-Jacobi function at zeroth and linear order in boundary fluctuations. We begin by evaluating \eqref{reggepointpart} in its background solution \eqref{deficitangle}. The bulk term is trivially zero for all edges of the triangulation except along the worldline $\mathcal{WL}$. On $\mathcal{WL}$, the Einstein-Hilbert term cancels the massive worldline term in \eqref{reggepointpart}. Consequently, at zeroth order, only the boundary term survives. This simplifies to a boundary term involving the boundary deficit angle $\omega_e$. Recalling its definition \eqref{deficitbound} and using \eqref{thetab}, we find
\begin{equation}
    \omega_e=\frac{2\pi\mu}{N_A}.\label{thetaevalue}
\end{equation}
Substituting this value in the boundary term of the Regge action \eqref{reggepointpart}, and noting that $N_T\times T=\beta$ we get 
\begin{equation}
    S_p^{(0)}=-\frac{\beta}{4G}\mu.\label{szero}
\end{equation}
For the first-order term, we need to take the first variation of the action, $\delta_{l_e}S_p$, and evaluate it in the background solution. By definition, the solution to the equations of motion satisfies $\delta_{l_e}S_p\big|_{\text{sol}} = 0$. This yields the Hamilton-Jacobi function at linear order in boundary length fluctuations,
\begin{equation}
    S_p^{(1)}=-\frac{1}{8\pi G }\frac{2\pi\mu}{N_A}\sum_{s,n}\tau(s,n).\label{action1bound}
\end{equation}
Now that we have derived the Hamilton-Jacobi function, we can proceed to analyze the second-order part of the action and the corresponding Hessian. This analysis will be crucial for understanding the role of diffeomorphism invariance discussed earlier


\subsection{Hessian and null vectors}
As discussed before, the second-order order contribution to the linearized Regge action is the Hessian \eqref{generalhessian}, which we compute in this section. Given its scaling dimensions, it will be useful to define the dimensionless hessian matrix $M_{ee'}$ in the following form
\begin{equation}
    H_{ee'}=\frac{L_eL_{e'}}{6V_\sigma}M_{ee'},
\end{equation}
where $L_e$ and $L_{e'}$ are the background lengths of the edges $e$ and $e'$ and $V_\sigma$ is the background volume of a tetrahedron \eqref{eq:vol_tetrahedron}.
Additionally, we introduce the rescaled fluctuation variables
\begin{equation}
    \hat{\lambda}_e:=\frac{L_e}{\sqrt{6V_\sigma}}\lambda_e,\label{rescaledlambda}
\end{equation}
where $\lambda_e$ denotes the length fluctuations \eqref{eq:length_fluc_s_n}.
Given the setup of the twisted thermal torus, it is convenient to Fourier transform the fluctuation variables \eqref{rescaledlambda} in both temporal and angular directions. These are given by
\begin{equation}
    \hat{\lambda}(k,n)=\frac{1}{\sqrt{N_A}}\sum_s e^{-i\frac{2\pi}{N_A}k\cdot s}\hat{\lambda}(s,n),\label{fourierangular}
\end{equation}
\begin{equation}
    \hat{\lambda}(k,\nu)=\frac{1}{\sqrt{N_T}}\sum_n e^{-i\frac{2\pi}{N_T}\left(\nu -\frac{\gamma}{2\pi}k\right)\cdot n }\hat{\lambda}(k,n),\label{fouriertemporal}
\end{equation}
where $k\in\{0,1,...,N_A-1\}$ and $\nu\in\{0,1,...,N_T-1\}$. The parameter $\gamma$ in \eqref{fouriertemporal} is called the twist angle and it measures the angular rotation made before identifying  $t\sim t+\beta$. It is given by
\begin{equation}
    \gamma=2\pi\frac{N_A}{N_T}.
\end{equation}
Now, we can write the second order term in \eqref{slinearized} in terms of these variables
\begin{equation}
    S_p^{(2)}=\frac{1}{16\pi G}\sum_{k,\nu}(\hat{\lambda}(k,\nu))^t\cdot\Tilde{M}(k,\nu)\cdot(\hat{\lambda}(-k,-\nu))\label{s2hessian},
\end{equation}
with
\begin{equation}
    (\hat{\lambda}(k,\nu))^t=(\hat{t}(\nu),\hat{r}(k,\nu),\hat{d}(k,\nu),\hat{\tau}(k,\nu),\hat{\alpha}(k,\nu),\hat{\eta}(k,\nu))
\end{equation}
the Fourier transform of the rescaled length fluctuations. 
As mentioned earlier, the form of the Hessian matrix remains unchanged compared with the vacuum case because the worldline part of \eqref{reggepointpart} is linear in the length variables. This was computed in \cite{Bianca:2015} and is given by
\begin{equation}
\Tilde{M}(k,\nu)=\left(
\scalebox{0.8}{$\begin{array}{cccccc}
 0 & -2x\sqrt{N_A}\delta_{k,0} & 0 & 0 & \sqrt{N_A}\delta_{k,0} & 0 \\
 -2x\sqrt{N_A}\delta_{k,0} & \Delta _k & 2 x \left(1-\omega _{v }\right)-\Delta _k & (\omega _k-1+2x)\omega _{v } & \omega _k \omega _{v }-1 & \omega _{v }-\omega _k \omega _{v } \\
 0 & 2 x \left(1-\frac{1}{\omega _{v }}\right)-\Delta _k & \Delta _k & \frac{1}{\omega _k}-1 & \frac{1}{\omega _{v }}-\omega _k & \omega _k-1 \\
 0 & \frac{1}{\omega_v}\left(\frac{1}{\omega_k}-1+2x\right) & \omega _k-1 & 1 & \frac{\omega _k}{2}-\frac{1}{2 \omega _{v }} & -\omega _k \\
 \sqrt{N_A}\delta_{k,0} & \frac{1}{\omega _k \omega _{v }}-1 & \omega _{v }-\frac{1}{\omega _k} & \frac{1}{2 \omega _k}-\frac{\omega _{v }}{2} & 1 & -\frac{\omega _{v }}{2}-\frac{1}{2} \\
 0  & \frac{1}{\omega _{v }}-\frac{1}{\omega _k \omega _{v }} & \frac{1}{\omega _k}-1 & -\frac{1}{\omega _k} & -\frac{1}{2 \omega _{v }}-\frac{1}{2} & 1 \\
\end{array}$},
\right)    \label{hessian}
\end{equation}
where
\begin{equation}
\label{omegafunction}
    \begin{split}
        \omega_k & := e^{i\frac{2\pi}{N_A}k}\hspace{2mm}, \hspace{10mm} \Delta_k :=2-\omega_k-\omega^{-1}_k,\\
        \omega_v & := e^{i\frac{2\pi}{N_T}v}\hspace{2mm}, \hspace{10mm} \Delta_v :=2-\omega_v-\omega^{-1}_v.
    \end{split}
\end{equation}
and $ v:=\nu -\frac{\gamma}{2\pi}k$. As discussed in \cite{Bianca:2015}, $\Delta_k$ and $\Delta_v$ are the eigenvalues of the Laplacians in the angular and temporal directions, respectively. \par
In order to identify the symmetries of the action, we need to compute the null eigenvectors of the matrix \eqref{hessian}. To begin, we find that 
\begin{equation}
    (n_\tau)^t(k\neq 0,\nu)=(0,0,1,1-\omega_v,0,1-\omega_v\omega_k),
\end{equation}
is a null vector that corresponds to the rescaling of time slices in triangulation. This is making the $n$-th time slice larger and compensating this by making the $(n-1)$-th slice smaller.
Additionally, there is a null vector that captures the global rescaling symmetry of the Hessian \eqref{generalhessian}. This is given by
\begin{equation}
    (n_{sc})^t(k=0,\nu=0)=\left(\frac{T^2}{\sqrt{N_A}},R^2,R^2+T^2,T^2,A^2,A^2+T^2\right).
\end{equation}
It is important to emphasize that these null vectors do not correspond to gauge symmetries of the bulk part of the path integral \eqref{formofpartfunction}. This distinction arises because the degrees of freedom we aim to integrate explicitly are those in the bulk. The two null vectors identified here are associated with boundary vertex displacements and global rescaling of the triangulation. In other words, these two null vectors do not affect the one-loop determinant of the bulk Hessian.  Null vectors that reside entirely in the bulk will be analyzed in the next section.

\subsection{Gauge modes} \label{gaugemodes}

Since the one-loop determinant is proportional to the inverse of the product of eigenvalues of the bulk Hessian, it is crucial to identify its null vectors. To identify these gauge modes, we analyze the bulk part of the Hessian, which is the upper-left $3\times3$ sub-matrix of \eqref{hessian}. For $k=0$, this matrix is given by 
\begin{equation}
    \Tilde{M}_\mathrm{bulk}(0,\nu):=\left(
\begin{array}{ccc}
 0 & -2x\sqrt{N_A} & 0 \\
 -2x\sqrt{N_A} & 0 & 2x(1-\omega_v) \\
 0 & 2x(1-\omega^{-1}_v) & 0\\ 
\end{array}
\right),\label{mbulkzero}
\end{equation}
which has a null vector given by
\begin{equation}
     n_t(k=0, \nu)=\left(\frac{(1-\omega_v)}{\sqrt{N_A}},0,-\omega_v,0,0,0\right).
\end{equation}
This corresponds to variations of the lengths in the temporal ($t$) and diagonal ($d$) directions.

With the bulk Hessian and its null vector identified, we can now substitute these into \eqref{s2hessian} to compute their contribution to the partition function. This involves integrating out the bulk variables $(\hat{r}, \hat{d})$ by diagonalizing \eqref{mbulkzero} and performing the Gaussian integral
\begin{equation}
        \int d\hat{r}(\nu)d\hat{d}(\nu)d\hat{r}(-\nu)d\hat{d}(-\nu)\exp{\left(-\frac{1}{2}(\hat{t},\hat{r},\hat{d})\cdot \Tilde{M}_\mathrm{bulk}(0,\nu)\cdot (\hat{t},\hat{r},\hat{d})^t\right)}
        =\frac{4\pi^2}{4x^2\Delta_v}.
\end{equation}
Since \eqref{mbulkzero} has a null vector, we have to take it into account in the gauge measure \eqref{singulargaugemeasure}. Doing so and taking the product over all the values of $\nu$ we get the following contribution to the one-loop partition function 
\begin{equation}
    (8\pi G\hbar)^{\frac{3}{2}N_T}\left(\frac{2\pi}{2x}\right)^{N_T}\frac{(6V_\sigma)^{\frac{3}{2}N_T}}{R^{2N_T}T^{N_T} }.\label{gaugemodecontribution}
\end{equation}
Geometrically, this null vector corresponds to invariance under displacements of the bulk vertices in the time direction. This symmetry extends to the point particle term of the action, as it reflects the parametrization invariance of the worldline.

For $k \neq 0$, the bulk Hessian does not have any null vectors. This stands in contrast to the vacuum case \cite{Bianca:2015}, where the $k = \pm 1$ modes also correspond to gauge modes. In the vacuum case, the $k = \pm 1$ null vectors reflected the invariance of the partition function under vertex displacements at the center of the torus triangulation. However, in the massive case, the center of the torus is no longer a gauge choice but represents the physical trajectory of a point particle. Geometrically, this corresponds to the particle's trajectory breaking a diffeomorphism symmetry that existed in the vacuum case. These remaining degrees of freedom, which are not associated with gauge symmetries, are referred to as physical modes.


\subsection{Physical modes}
Now, to compute the partition function contribution from the physical modes, we need to consider the full Hessian \eqref{hessian}. First, we integrate out the $\hat{d}$ variables by solving the equations $\frac{\delta S}{\delta \hat{d}(k,\nu)} = 0$ and $\frac{\delta S}{\delta \hat{d}(-k,-\nu)} = 0$ for $\hat{d}(k,\nu)$ and $\hat{d}(-k,-\nu)$. After finding these solutions, we substitute them into \eqref{hessian} and extract the coefficients of the resulting matrix
\begin{equation}
   \Tilde{M}_r(k,\nu)= \left(
\scalebox{0.7}{$\begin{array}{cccc}
 2 x \Delta _{v } \left(1-\frac{2 x}{\Delta _k}\right) & \frac{\left(\Delta _k-2 x\right) \left(\frac{1}{\omega _k}-\omega _{v }\right)}{1-\frac{1}{\omega _k}} & \left(1-\frac{1}{\omega _{v}}\right) \left(\omega _{v} \omega _k-1\right) \left(1-\frac{2 x}{\Delta _k}\right) & \frac{\left(\Delta _k-2 x\right) \left(\omega _{v }-1\right)}{1-\frac{1}{\omega _k}} \\
 \frac{\left(\Delta _k-2 x\right) \left(\omega _k-\frac{1}{\omega _{v }}\right)}{1-\omega _k} & 0 & -\frac{\left(\omega _k+1\right) \left(\omega _k \omega _{v }-1\right)}{2 \left(\omega _k-1\right) \omega _{v }} & 0 \\
 \left(1-\omega _{v }\right) \left(\frac{1}{\omega _{v } \omega _k}-1\right) \left(1-\frac{2 x}{\Delta _k}\right) & -\frac{\left(\omega _k+1\right) \left(\omega _k \omega _{v }-1\right)}{2 \left(\omega _k-1\right) \omega _k} & \frac{\omega _{v } \omega _k-\omega _k+\frac{1}{\omega _k \omega _{v }}-\frac{1}{\omega _k}}{\Delta _k} & \frac{\left(\omega _k+1\right) \left(\omega _{v }-1\right)}{2 \left(\omega _k-1\right)} \\
 \frac{\left(\Delta _k-2 x\right) \left(\frac{1}{\omega _{v }}-1\right)}{1-\omega _k} & 0 & \frac{\left(\omega _k+1\right) \left(\omega _{v }-1\right)}{2 \left(\omega _k-1\right) \omega _{v }} & 0 \\ \end{array}$}
\right).\label{raction}
\end{equation}
Recall that in this step we are computing Gaussian integrals and summing over all the modes $(k,\nu)$ with $|k|>0$. The factor we get from integrating out the $\hat{d}$ variables is
\begin{equation}
\begin{aligned}
\prod_{k=1}^{N_A-1} & \prod_{\nu=1}^{N_T-1} \int d\hat{d}(k,\nu) d\hat{d}(-k,-\nu) \exp \left( -\frac{1}{2 \times 8 \pi G \hbar}\hat{d}(k,\nu) \cdot \Delta_k \cdot  \hat{d}(-k,-\nu)\right)\\
      & = (8 \pi G \hbar)^{\frac{N_T(N_A-1)}{2}}(2 \pi)^{\frac{N_T(N_A-1)}{2}}N_A^{-N_T}.\label{physicalmodecontribution}
\end{aligned}
\end{equation}


\subsection{One-loop determinant}
The one-loop determinant, by definition, is the determinant of the Hessian restricted to the bulk variables. Since there are no zero eigenvalues for $k\neq 0$, we consider the upper-left part of \eqref{hessian}
\begin{equation}
    \Tilde{M}_\mathrm{bulk}(k\neq 0,\nu):=\left(
\begin{array}{cc}
 \Delta _k & 2 x \left(1-\omega _{v }\right)-\Delta _k \\
 2 x \left(1-\omega^{-1}_{v}\right)-\Delta _k & \Delta _k \\
\end{array}
\right).\label{hess1loop}
\end{equation}
From \eqref{raction}, we observe that integrating out the $\hat{r}$ variables corresponds to a Gaussian integral with the following exponent
\begin{equation}
    2x\Delta_v\left(1-\frac{2x}{\Delta_k}\right).
\end{equation}
Next, by taking the product over all the modes $(k\neq 0,\nu)$, we obtain
\begin{equation}
    (2x)^{N_T(N_A-1)}\left(\prod_{k=1}^{N_A-1}\left(1-\frac{2x}{\Delta_k}\right)^{N_T}\right)\left(\prod_{k=1}^{N_A-1}\prod_{\nu=0}^{N_T-1}\Delta_v\right).
\end{equation}
Using the definition of $\Delta_v$ and adding the corresponding factors of $\pi$ and $G\hbar$ from the action, we arrive to the following expression 
\begin{equation}
    (8\pi G\hbar)^{\frac{N_T(N_A-1)}{2}}(2\pi)^{\frac{N_T(N_A-1)}{2}}(2x)^{\frac{-N_T(N_A-1)}{2}}f(x,N_A)^{-\frac{N_T}{2}}\left[\prod_{k=1}^{N_A-1}(2-2\hspace{1mm}\mathrm{cos}(\gamma k))\right]^{-\frac{1}{2}},\label{oneloopcontribution}
\end{equation}
where
\begin{equation}
\label{pochhammer}
    f(x,N_A):=\prod _{k=1}^{N_A-1} \left(1-\frac{2 x}{\Delta_k}\right)=
    \frac{ \left(\frac{1}{1-x+\sqrt{(x-2) x}};e^{\frac{i2\pi i}{N_A}}\right)_{N_A} \left(-\frac{1}{x-1+\sqrt{(x-2) x}};e^{\frac{i2\pi i}{N_A}}\right)_{N_A}}{2 \left(\left(e^{\frac{i2\pi i}{N_A}};e^{\frac{i2\pi i}{N_A}}\right){}_{N_A-1}\right)^2},
\end{equation}
and $(\cdot ; \cdot)_N$ is the $q$-Pochhammer symbol.\par
We combine the integration contributions from the gauge modes \eqref{gaugemodecontribution}, the measure, and the physical modes \eqref{physicalmodecontribution} along with the one-loop determinant $D$ \eqref{oneloopcontribution} in \eqref{formofpartfunction}. The resulting expression is
\begin{equation}
    \begin{split}
        D=2^{-N_T}(2\pi)^{-\frac{N_TN_A}{2}}\left(\frac{R}{A}\right)^{N_T(N_A-1)}(ART)^{-\frac{N_TN_A}{2}}(A^2T)^{N_T}\left(4-\frac{A^2}{R^2}\right)^{-N_T\left(\frac{N_A}{4}+1\right)}\\
        \left(\frac{N_A^2}{2^{N_A-2}}f(x,N_A)\right)^{-\frac{N_T}{2}}\left[\prod_{e\in bdry}\frac{L_e^{1/2}}{(8\pi G\hbar)^{1/4}}\right]\left[\prod_{k=1}^{(N_A-1)/2}\frac{1}{|1-q^k|^2}\right]\\
        := \tilde{D}\prod_{k=1}^{(N_A-1)/2}\frac{1}{|1-q^k|^2},
    \end{split}
\end{equation}
where we denote $q=\exp(i\gamma)$. Notably, $\gamma$ plays a role analogue to the modular parameter in the AdS$_3$ case.

Using this result and the zeroth and first order parts of the action, \eqref{szero} and \eqref{action1bound} respectively, we can write the resulting partition function
\begin{equation}
\boxed{
    Z(\{\tau(s,n),\alpha(s,n),\eta(s,n)\})=\exp{\left(\frac{1}{\hbar}\frac{\beta}{4G}\mu\right)}\tilde{D} \prod_{k=1}^{(N_A-1)/2}\frac{1}{|1-q^k|^2}\hspace{1mm}e^{-F(\{\tau(s,n),\alpha(s,n),\eta(s,n)\})}.\label{resultingZ}
    }
\end{equation}
Let us analyze each component of this expression in detail.

\begin{enumerate}
    \item \textbf{The constant \(\tilde{D}\):}  
    This term depends on the background triangulation through the lengths of the building blocks and the number of segments along the radial and time directions. While it also depends on \(G\) and \(\hbar\), these dependencies are only through power-law factors, which do not contribute to the BMS$_3$ character. As a result, \(\tilde{D}\) is inherently discretization-dependent. Importantly, we expect it to approach a constant value in the continuum limit, which has been confirmed through numerical checks.

    \item \textbf{The product \(|1 - q^k|^{-2}\):}  
    Similar to the vacuum case \cite{Bianca:2015}, this factor becomes singular when \(\gamma \times k\) is an integer multiple of \(2\pi\). In the discrete geometry case, this is not problematic because \(\gamma\) takes rational values and \(k\) is an integer. However, in the continuum limit, these singularities can arise in general. To address this, one would need to regularize by taking \(\gamma \rightarrow \gamma + \epsilon\). We will further comment on this issue when performing the computation directly in the continuum.

    \item \textbf{Agreement with the massive BMS$_3$ character:}  
    In the refinement limit, \(N_A \to \infty\), \eqref{resultingZ} matches the character of a massive BMS$_3$ particle \eqref{bms3massive}. By comparing the two expressions and setting \(\alpha^0 = i\beta\), we identify the following relations
\begin{equation}
\label{masscharge}
    m=\frac{M}{\hbar}\hspace{2mm},\hspace{2mm}c=\frac{6}{G\hbar}.
\end{equation}
    These relations establish a correspondence between the geometrical mass \(M\) and the BMS$_3$ mass \(m\), as well as between Newton's constant and the central charge \(c\). The latter is consistent with the value reported in the literature \cite{Oblak:2015BMS}.
\end{enumerate}
This result provides a clear geometric interpretation of how diffeomorphism symmetry breaking distinguishes the vacuum and massive cases within the discrete geometry framework. Specifically, the difference is encoded in the value of
$k$ where the product expansion of the character starts: 
$k=2$ for the vacuum and $k=1$ for the massive case. This shift reflects the reduction in diffeomorphism invariance due to the presence of the massive particle. In the vacuum, full diffeomorphism invariance allows for gauge-fixing the position of the torus center arbitrarily. In contrast, when a massive particle is present, its worldline anchors the geometry, breaking this gauge freedom and fixing the center physically. Thus, the appearance of the additional 
$k=1$ contributions in the massive BMS$_3$ character is a direct manifestation of diffeomorphism symmetry breaking in Regge gravity.


Now that we have integrated the bulk degrees of freedom of the system, we turn our attention to the analysis of the resulting boundary degrees of freedom contained in the effective boundary action  $F$ in \eqref{resultingZ}.

\subsection{Effective boundary action}
The resulting effective boundary action  $F$ in \eqref{resultingZ} corresponds to the linear and quadratic parts in boundary fluctuations of \eqref{slinearized}. The linear contribution is given by \eqref{action1bound}, while the quadratic part is obtained by integrating out the bulk fluctuations of the Hessian \eqref{hessian}. For $k\neq 0$, this integration involves the variables 
$\hat{d}$ and $\hat{r}$ from \eqref{hessian}. The resulting boundary matrix is
\begin{equation}
\begin{split}
    \Tilde{M}_\mathrm{b}(k,\nu)=\frac{1}{2x}\left(
\begin{array}{ccc}
 \frac{-\Delta_k\Delta_{kv}}{\Delta_v} &  \frac{-\Delta_{kv}(1-\omega_k)}{(1-\omega_v)} &  \frac{-\Delta_{k}(\omega_k-\omega_v^{-1})}{(1-\omega_v^{-1})} \\
 \dots & -\Delta_{kv} & (1-\omega_k^{-1})(1-\omega_k\omega_v) \\
 \dots & \dots & -\Delta_k\\ 
\end{array}
\right) \\
+\left(
\begin{array}{ccc}
 \frac{\Delta_{kv}}{\Delta_v} &  \frac{-(1+\omega_v)(\omega_v^{-1}-\omega_k)}{2(1-\omega_v)} &  -\frac{(\omega_k-\omega_v^{-1})}{(1-\omega_v^{-1})} \\
 \dots & 1 & -\frac{1}{2}(1+\omega_v) \\
 \dots & \dots & 1\\ 
\end{array}
\right),
\end{split}
\end{equation}
where the missing entries are such that the matrix is hermitian and we define
\begin{equation}
    \Delta_{kv}=2-\omega_k\omega_v-(\omega_k\omega_v)^{-1}.
\end{equation}
Note that, while the coefficients are identical to those in the vacuum case \cite{Bianca:2015}, the presence of the mass is implicitly encoded in the definition of $x$ as given in \eqref{definitionx}. This matrix makes up for the second-order contribution to the action that has the following form
\begin{equation}
    8\pi G S_b^{(2)}=-\frac{1}{2}\sum_{k,\nu}(\hat{\lambda}_b(k,\nu))^t\cdot \Tilde{M}_b(k,\nu)\cdot (\hat{\lambda}_b(k,\nu)),\label{eq:second_ord_bdy_action}
\end{equation}
where $\hat{\lambda}_b(k,\nu)=(\hat{\tau}(k,\nu),\hat{\alpha}(k,\nu),\hat{\eta}(k,\nu))$.\par
In order to facilitate comparison with the continuum case, is it useful to change variables from length fluctuations to metric fluctuations. By following the procedure outlined in \cite{Bianca:2015}, we find
\begin{equation}
    \begin{split}
        (h_{ab}+\delta h_{ab})e^a_\tau e^b_\tau=(T+\tau)^2=T^2+2T\tau+\mathcal{O}(\tau^2),\\
        (h_{ab}+\delta h_{ab})e^a_\alpha e^b_\alpha=(A+\alpha)^2=A^2+2A\alpha+\mathcal{O}(\alpha^2),\\
        (h_{ab}+\delta h_{ab})(e^a_\tau+ e^a_\alpha)(e^b_\tau+ e^b_\alpha)=(\sqrt{A^2+T^2}+\eta^2)^2\\
        =A^2+T^2+2\sqrt{A^2+T^2}\eta+\mathcal{O}(\eta^2),
    \end{split}
\end{equation}
where $h_{ab}$ and $\delta h_{ab}$ are the background boundary metric and its fluctuation respectively, and $e^a=(e^a_\tau, e^a_\alpha)$ is a unit vector in the tangent space of the boundary of the twisted thermal torus. These equations fix the background boundary metric to be $h_{ab}=diag(T^2,A^2)$. Incorporating the definition \eqref{rescaledlambda}, we have
\begin{equation}\label{eq:tometricvariables}
\left(
    \begin{array}{c}
        \delta h_{\tau\tau}  \\
         \delta h_{\alpha\alpha} \\
         \delta h_{\tau\alpha}
    \end{array}\right)=
    \sqrt{6V_\sigma}
    \left(\begin{array}{ccc}
        2 & 0 & 0 \\
        0 & 2 & 0  \\
        -1 & -1 & 1 
    \end{array}\right)
    \left(
    \begin{array}{c}
        \hat{\tau}  \\
         \hat{\alpha} \\
         \hat{\eta}
    \end{array}\right)+\mathcal{O}(\lambda_b^2),
\end{equation}
where $\mathcal{O}(\lambda_b^2)$ denotes higher order terms in boundary length fluctuations. The linear order boundary length fluctuations are useful for writing the second-order part of the boundary action \eqref{eq:second_ord_bdy_action}. However, there is the first-order part of the action, \eqref{action1bound}, which is linear in boundary length fluctuations, $\tau$. To obtain the quadratic order contribution from the first-order action, we must also express $\tau$ in terms of metric variables up to second order. This is given by
\begin{equation}
    \tau=\frac{1}{2T}\delta h_{\tau\tau}-\frac{1}{8T^3}(\delta h_{\tau\tau})^2+\mathcal{O}((\delta h_{\tau\tau})^3).
\end{equation}
Finally, by combining the second-order action up to linear order and the first-order action up to quadratic order, we obtain the following matrix
\begin{equation}
\begin{split}
    \Tilde{M}^h_\mathrm{b}(k,\nu)=-\frac{1}{8(6V_\sigma)}\left(\frac{1}{x}\left(
\scalebox{0.8}{$\begin{array}{ccc}
 \frac{-\Delta_k^2}{\Delta_v} & \Delta_{k} &  2\frac{\Delta_{k}(1-\omega_k)}{(1-\omega_v^{-1})} \\
 \dots & \Delta_{v} & 2(1-\omega_k)(1-\omega_v) \\
 \dots & \dots & 4\Delta_k\\ 
\end{array}$}
\right)-\left(
\scalebox{0.8}{$\begin{array}{ccc}
 \frac{2\Delta_{k}}{\Delta_v} &  (1-\omega_k) &  4\frac{(1-\omega_k)}{(1-\omega_v^{-1})} \\
 \dots & \Delta_v & 2(1-\omega_v) \\
 \dots & \dots & 8\\ 
\end{array}$}
\right) \right)\\ +\frac{\pi\mu}{8N_AT^3}\left(
\scalebox{0.8}{$\begin{array}{ccc}
 1 &  0 &  0 \\
 \dots & 0 & 0 \\
 \dots & \dots & 0\\ 
\end{array}$}
\right).
\end{split}\label{hessboundmetric}
\end{equation}
The effective boundary action in metric fluctuations is 
\begin{equation}\label{eq:S2_bound_Regge}
    {S^h_p}^{(2)}=-\frac{1}{16\pi G}\sum_{k,\nu}(\delta h(k,\nu))^t\cdot \Tilde{M}^h_\mathrm{b}(k,\nu)\cdot (\delta h(k,\nu)).
\end{equation}
which incorporates both the boundary Hessian after integrating bulk degrees of freedom as well as the first order part \eqref{action1bound}.\par
We will come back to this effective boundary action to take the continuum limit of the discrete geometry model. This step is essential for comparing the outcomes of the Regge gravity computations with the results of the continuum theory in Section \ref{section: Continuum}.


\subsection{Comparison with the vacuum case}\label{section: comp_w_vacuum}

In this section, we compare our results with those of the vacuum case reported in \cite{Bianca:2015}. This serves as both a valuable consistency check and a means to identify the remnants of diffeomorphism breaking introduced by the presence of the massive particle.


In the pure gravity case, the $k = \pm1$ modes were identified as gauge modes. However, in the presence of a massive particle at the center of the torus, these modes become physical. This change is directly linked to the breaking of radial diffeomorphism invariance caused by the particle's worldline. As a consistency check, we track the deformation of the eigenvalues of \eqref{hess1loop} in the limit $M \ll 1$ (or equivalently in the vanishing deficit angle limit $\mu \sim 1$) and prove that the $k = \pm1$ eigenvalue vanishes for $M=0$. We begin by expanding $x$, given by \eqref{definitionx}, around $M=0$ in the following way
\begin{equation}
    x=1-\mathrm{cos}\left(\frac{2\pi}{N_A} \right)-\frac{8\pi GM}{N_A}\mathrm{sin}\left(\frac{2\pi}{N_A} \right)+\mathcal{O}(M^2).\label{xexpansion}
\end{equation}
This implies 
\begin{equation}
    \Delta_{k=\pm 1}=2(x+aM)+\mathcal{O}(M^2)\label{delta1expansion},
\end{equation}
where $a=\frac{8\pi G}{N_A}\mathrm{sin}\left(\frac{2\pi}{N_A} \right)$. With this expansion, we can rewrite \eqref{hess1loop} in a more explicit form, which facilitates the comparison between the massive and vacuum cases
\begin{equation}
    \Tilde{M}_\mathrm{bulk}(\pm 1,\nu)\approx\left(
\begin{array}{cc}
 x+aM & -aM-x\omega _v\\
 -aM-x\omega^{-1}_v & x+aM \\
\end{array}
\right).
\end{equation}
Next, we compute the eigenvalues of this bulk Hessian, which are given by
\begin{equation}
    \lambda^1_{\pm}=(x+aM)\pm\sqrt{(x+aM)^2-a\Delta_vM x}.
\end{equation}
Expanding these expressions around $M=0$, we find
\begin{equation}
\lambda^1_\pm\approx\left\{
\begin{aligned}
 a\Delta_vM,\\
2x-a\Delta_vM.
\end{aligned}
\right.
\label{k1eigenvalues}
\end{equation}
These eigenvalues reduce to those reported in the vacuum case \cite{Bianca:2015} when $M=0$. Furthermore, the eigenvalue $\lambda^1_+=a\Delta_vM$, which corresponds vanishes in the vacuum case, is associated with the eigenvector
\begin{equation}
    n_r(k=\pm 1, \nu)\approx\left(0,\omega_v,1,0,0,0\right)+\frac{aM}{2x}(1-\omega_v^2)\left(0,1,0,0,0,0\right).
\end{equation}
This eigenvector also reduces to the vacuum case null eigenvector in the $M\rightarrow0$ limit, providing a consistent result.




\section{Partition function in the continuum}\label{section: Continuum} 

In this section, we extend the discrete geometry computations to the continuum setting, focusing on how the mass modifies the boundary action \eqref{discreteXr} and its relation to the massive BMS$_3$ character. Following the methods in \cite{asante2019holographic}, we introduce a massive point particle at the center of the twisted thermal flat torus and derive the corresponding dual boundary field action.

We begin by setting up the geometric framework using Gaussian coordinates. This choice naturally foliates the bulk into constant-radius surfaces, simplifying the integration over bulk variables. Within this setup, we parametrize metric perturbations in terms of diffeomorphism-generating vector fields. The Hamilton-Jacobi functional can then be conveniently expressed in terms of these vector fields, which generate on-shell metric perturbations. Since these vector fields are non-local on the boundary metric, this suggests the existence of a quasi-local dual boundary field theory.

Similarly to the vacuum case, we propose a Liouville-like boundary theory that reproduces the Hamilton-Jacobi functional for 3D gravity. We can do this because the presence of a test particle introduces a conical defect that modifies the spacetime globally, while the local structure remains identical to the vacuum case. Consequently, the Hamilton-Jacobi functional for the action \eqref{actionpointpart} coincides with that of vacuum gravity. Moreover, the equations of motion impose a key condition: the scalar field must correspond to the geodesic length from the center of the torus to its boundary.

However, as noted in \cite{asante2019holographic}, the geodesic length alone is a degenerate observable, meaning that multiple Liouville-like boundary theories can yield the same scalar field. To resolve this, \cite{asante2019holographic} introduced a smoothness condition at the central axis of the torus, ensuring that the spacetime remains free of singularities at 
$r=0$. In our work, we show that this smoothness condition acquires a different physical meaning in the presence of a point particle: the central axis now coincides with the particle's worldline, turning what was previously a gauge choice into a physical feature of the geometry.

This distinction is a direct reflects the breaking of diffeomorphism symmetry by the point particle. Unlike in pure gravity, where diffeomorphisms can smooth out local features everywhere, the particle explicitly breaks radial diffeomorphism invariance, constraining the allowed deformations of the geometry. As a result, the smoothness conditions are now dictated by the physical presence of the particle itself.

This key difference extends and reinterprets the results of \cite{asante2019holographic}: in the massive case, the modification of the mode expansion and boundary conditions arises naturally from symmetry breaking, rather than from additional gauge-fixing. This geometric mechanism precisely mirrors our findings in the discrete Regge setting, reinforcing the robustness of our results.

\subsection{Background metric and perturbations} \label{Assumptions and Conventions} 

We begin with the same background geometry as in the discrete case: the twisted thermal flat torus. To express its metric, we introduce Gaussian coordinates, writing it as
\begin{equation}
\label{gaussian}
g_{ab}dx^adx^b = dr^2 + h_{AB}dy^Ady^B,
\end{equation}
where $a, b, \dots$ denote spacetime indices, and $A, B, \dots$ label spatial coordinates on constant-$r$ surfaces. In these coordinates, the extrinsic curvature $K_{AB} = \frac{1}{2} \partial_\perp h_{AB}$ for a constant-$r$ surface reads
\begin{equation}
\begin{aligned}
\label{eqn:twistedflat1}
    K_{AB} = & r \delta^\theta_A \delta^\theta_B.
\end{aligned}
\end{equation}
\begin{figure}[h]
  \centering
    \includegraphics[width=0.7\textwidth]{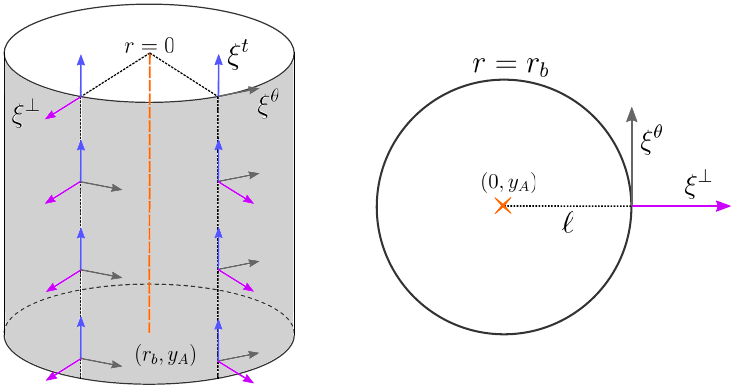}
         \put(-250,-10){a)}
     \put(-100,-10){b)}
  \caption{a) Diffeomorphism-generating vector fields on the Gaussian surfaces of constant radii. b) A constant time slice and the geodesic length $\ell$ from the center of the torus to the boundary. The point particle's worldline is shown in orange.}  \label{Fig:continuum}
\end{figure}\\
To account for metric fluctuations $\gamma_{ab}$ around this background geometry, we write the full metric as
\begin{equation}
g_{ab}^{\text{full}} = g_{ab} + \gamma_{ab}.
\end{equation}
For convenience, we denote the components of the metric perturbation at constant background radii surfaces as $\gamma_{\perp \perp}$, $\gamma_{\perp A}$, and $\gamma_{AB}$.

Since we aim to replicate the results obtained in the discrete setting, our goal is to compute the Hamilton-Jacobi functional for linearized gravity coupled to a point particle in the continuum. However, introducing a point particle creates a conical defect in the background spacetime, making it homogeneous everywhere except at the center of the torus. This is reflected in the equations of motion \eqref{eqnofmotion1}, which require the Ricci scalar to be nonzero at the center. Consequently, the equations of motion are not diffeomorphism equivalent to a homogeneously curved spacetime.

Despite this, the discrete geometry computations provide a key guiding principle: the only way in which the point particle affects the theory’s diffeomorphisms is by breaking invariance in the radial direction. To verify whether this remains true in the continuum setting, we follow \cite{asante2019holographic} and use diffeomorphism-generating vector fields to check whether the solutions display diffeomorphism invariance in the radial direction.

The relations between spacetime covariant derivatives and spatial covariant derivatives for a diffeomorphism-generating vector $\xi$ are
\begin{equation}
    \nabla_A \xi_B = D_A \xi_B + K_{AB} \xi_\perp,
\end{equation}
\begin{equation}
    \nabla_A \xi_\perp = D_A (\xi_\perp) + K^B_A \xi_B.
\end{equation}
Here, $\nabla$ denotes the covariant derivative with respect to metric $g$, $D$ is the covariant derivative with respect to $h$,
\begin{equation}
\label{eqn:twistedflat3}
\begin{aligned}
    D_{AB} =& -\frac{2}{r} h_{AB} \partial_t^2.
\end{aligned}
\end{equation}
Since $\xi_\perp$ is a vector perpendicular to the boundary surface, it is treated as a scalar with respect to the boundary coordinate system. This implies that $D_A \xi_\perp = \partial_A \xi_\perp$.

The metric $\gamma_{AB}$ can be parametrized using the vector components $\xi^\perp$ and $\xi^A$ in the following way
\begin{equation}
\label{eqn:liederivative}
\begin{aligned}
    \gamma_{AB} = [\Ll_\xi g]_{AB} = \nabla_A \xi_B + \nabla_B \xi_A
                = 2 \xi^\perp K_{AB} + [\Ll_{\xi^{||}} h]_{AB}.
\end{aligned}
\end{equation}
We assume the relation between the boundary metric perturbations and the diffeomorphism-generating vector fields is invertible. 
Using this, the vector components $\xi^\perp$ and $\xi^A$ are determined  by
\begin{equation}
\label{xi1}
\begin{aligned}
    \Delta \xi^\perp = & \Pi^{AB} \gamma_{AB}, \\
    D^A_B \xi^B = & 2 (K^{BC} - Kh^{BC}) \delta \prescript{2}{}{\Gamma^A_{BC}}
\end{aligned}
\end{equation}
where
\begin{equation}
\label{xi2}
\begin{aligned}
    \Delta = & 2(K^{CD} - Kh^{CD})D_CD_D - \prescript{2}{}{R} K =- 2r^{-1} \partial^2_t, \\
    D^A \hspace{1pt} _B = & 2 (K^{BC} - Kh^{CD})D_CD_Dh^A \hspace{1pt} _B - \prescript{2}{}{R} K^A \hspace{1pt} _B , \\
    \Pi^{AB} = & D^AD^B - h^{AB} D_CD^C - \frac{1}{2} \prescript{2}{}{R} h^{AB} , \\
    \delta  \prescript{2}{}{\Gamma^A_{BC}} = & \frac{1}{2} h^{AD} (D_B \gamma_{AC} + D_C \gamma_{BA} - D_A \gamma_{BC}) .
\end{aligned}
\end{equation}
As mentioned in \cite{asante2019holographic}, this implies that in order to solve for $\xi^\perp$ and $\xi^A$, the operators $\Delta$ and $D^A \hspace{1pt} _B$ need to be inverted. Thus, these fields are non-local functionals of the boundary metric. Furthermore, the previous expressions suggest a relationship between $\Pi^{AB} \gamma_{AB}$ and the first variation of the boundary Ricci scalar $\delta (\prescript{2}{}{R})$ given by,
\begin{equation}
\begin{aligned}
     \Pi^{AB} \gamma_{AB} = & \delta (\prescript{2}{}{R})
\end{aligned}
\end{equation}
This relation arises due to the invariance of $\xi^\perp$ under linearized boundary tangential diffeomorphisms, which leads to the vanishing of $\Pi^{AB}$ on perturbations induced by tangential diffeomorphisms. 

To finish, we write the lapse $\gamma_{\perp \perp}$ and shift $\gamma_{\perp A}$ of the metric perturbations as functions of the generating vector fields $(\xi^\perp , \xi^A)$ in the following way
\begin{equation}
\label{eqn:diffeometric0}
\begin{aligned}
    \gamma_{\perp \perp} = & 2 \partial_\perp \xi^\perp , \\
     \gamma_{\perp A}= & \nabla_\perp \xi_A + \nabla_A \xi_r 
     = D_A \xi^\perp + h_{AB} \partial_\perp \xi^B .
\end{aligned}
\end{equation}

\subsection{Hamilton-Jacobi functional} \label{Hamilton-Jacobi functional}
In this section, we examine the first- and second-order variations of the Hamilton-Jacobi functional for the action \eqref{actionpointpart}. We treat the point particle as a test particle and focus solely on variations with respect to the metric, neglecting variations with respect to the particle’s degrees of freedom. Since the point particle only introduces a conical defect, we take an alternative approach by considering a partition function over a flat spacetime with this defect. This follows the same procedure as in the vacuum case, with the crucial distinction that we will carefully account for the angular identification at a later stage. As we will see, this is sufficient to reproduce the massive BMS$_3$ character.

The first-order variation of the gravitational action \eqref{actionpointpart} is given by
\begin{equation}
\begin{aligned}
    \delta S = & -\frac{1}{16 \pi G} \int_{\mathcal M} d^3x \sqrt{g} \left( \frac{1}{2}R g^{ab} - R^{ab} \right) \delta g_{ab} - \frac{1}{16 \pi G} \int_{\partial \mathcal M} d^2y \sqrt{h} (Kh^{AB} - K^{AB}) \delta g_{AB} 
\end{aligned}
\end{equation}
We use the first-order variation of the action to determine the first-order on-shell action, the equations of motion evaluated on the background, and the momentum conjugated to the metric $\pi^{AB} = \sqrt{h} (K^{AB} - Kh^{AB})$. We use the parametrization $\delta g_{ab} = \gamma_{ab} = \Ll_\xi g_{ab}$ and the fact that, in Gaussian coordinates, the extrinsic curvature is given by $K_{AB} = \frac{1}{2} \partial_\perp h_{AB}$. Since $h_{AB}$ only varies for the boundary metric fluctuations, the first-order of the on-shell action evaluates to,
\begin{equation}
    S^{(1)}_{HJ} =  -\frac{1}{16 \pi G} \int_{\partial \mathcal M} d^2 y \sqrt{h} (\prescript{2}{}{R}) 
\end{equation}
This corresponds to the first-order Hamilton-Jacobi functional in the massless case \cite{asante2019holographic}, where the boundary Ricci scalar comes from the Gauss-Codazzi relations 
\begin{equation}
\label{gausscodazi}
    K^2 - K_{AB}K^{AB} = \prescript{2}{}{R} , \hspace{10pt} D_AK^A_B - D_BK = 0.
\end{equation}
The presence of first-order terms of $\xi$ will lead to higher-order terms necessary to make the effective action invariant under boundary tangential diffeomorphisms.

The second-order part of the Hamilton–Jacobi functional in $\xi^\perp$ evaluates to
\begin{equation}
\label{eqn:SHJVM}
\begin{aligned}
     S_{HJ}^{(2)} 
     = & -\frac{1}{32 \pi G} \int_{\partial\mathcal{M}} d^2 y \sqrt{h} (\xi^\perp \Delta \xi^\perp - \xi^A D_{AB} \xi^{B}) \  \\
\end{aligned}
\end{equation}
This means that the form of the second-order Hamilton-Jacobi functional unchanged with respect to the vacuum case \cite{asante2019holographic}\footnote{Another way of seeing this is by taking the variations of the full action \eqref{actionpointpart} with respect to the metric and noting that, since the worldline of the point particle is parametrization invariant in the time direction and the particle is spinless, the second-order part of the Hamilton-Jacobi functional quadratic in $\xi^\perp$ is proportional to $K_{tt}$ which vanishes.}.


\subsection{Geodesic length and boundary field} \label{geolength}
Since the form of the dual boundary field theory is not immediately evident from the form of \eqref{eqn:SHJVM}, in \cite{asante2019holographic}, the authors sought a boundary field action that satisfies the following criteria:
\begin{itemize}
\item It reproduces the equations of motion for the geodesic length.
\item It captures the boundary diffeomorphism-invariant part of the Hamilton-Jacobi functional.
\item Its one-loop determinant yields the vacuum BMS$_3$ character.
\end{itemize}
Their approach identifies the geodesic length spanning from the boundary to the central axis as the relevant boundary field. As noted in \cite{asante2019holographic}, this geodesic length serves as an observable characterizing the embedding of the boundary in the bulk solution. Moreover, in the discrete case, the triangulation structure closely resembles that of the vacuum case \cite{Bianca:2015}, as illustrated in Figures \ref{spatialslice} and \ref{discretization}, where edge lengths extend from the boundary to a central point or axis in the bulk. Motivated by this similarity, we adopt the same procedure in the continuum case.


We consider a geodesic from a point on the boundary $(r_b, y^A)$ to the central axis $(r_b=0, y^A)$. A geodesic affinely parametrized with respect to a background metric of the Gaussian form satisfies the geodesic equation 
\begin{equation}
    \frac{dx^a}{d\tau} \nabla_a \frac{dx^a}{d\tau} = \Gamma^b_{\perp \perp} (r_2 -r_1)^2 = 0,
\end{equation}
where $\tau$ is the affine parameter and $r_2 $, $r_1$ are fixed endpoints. Its length is given by
\begin{equation}
    \ell = \frac{1}{2} \int^{r_{2}}_{r_{1}} dr \ \gamma_{\perp \perp}
\end{equation}
at first order in metric perturbations.
Using \eqref{eqn:diffeometric0}, we have that 
\begin{equation}
    \ell = \xi^\perp(r_2) - \xi^\perp (r_1).
\end{equation}
This establishes the relation between the geodesic length, metric perturbations, and $\xi^\perp$. 


In order to obtain the effective action for geodesic length that displays the same diffeomorphism and gauge symmetries as in \cite{asante2019holographic}, we introduce a Lagrange multiplier $\lambda$ in the following form
\begin{equation}
\label{eqn:lagrangrmul}
\begin{aligned}
    -8 \pi G S^{(2)}_\lambda = & \frac{1}{4} \int_{\mathcal{M}} d^3x \sqrt{g} \gamma_{ab} (V^{abcd} \gamma_{cd} + \frac{1}{2} G^{abcdef} \nabla_c \nabla_d \gamma_{ef}) + \\
    & \frac{1}{4} \int_{\partial \mathcal{M}} d^2y \sqrt{h} \gamma_{ab} ((B_1)^{abcd} \gamma_{cd} + (B_2)^{abcde} \nabla_c \gamma_{de}) + \\
    & \frac{1}{4} \int_{(\partial \mathcal{M})_{out}} d^2y \lambda(y) (\rho(y) - \ell[\gamma_{\perp \perp}]).
\end{aligned}
\end{equation}
The first two terms correspond to the bulk and boundary contributions to the gravitational action, respectively. These terms are not uniquely determined, as their form can be modified through integration by parts. We choose them such that the bulk term vanishes on-shell (see Appendix A). The parameter $\lambda$ is treated as a first-order variable in the boundary metric perturbation and it is a scalar density with respect to the boundary metric. The field $\rho$ is a scalar defined on the boundary. 

Notably, there is no explicit particle term in \eqref{eqn:lagrangrmul}. This is because adding such a term is unnecessary: in the equations of motion derived from the action \eqref{actionpointpart}, the mass term (or the energy-momentum tensor) remains invariant under metric variations. Furthermore, the metric solution to \eqref{eqnofmotion1} corresponds to a twisted flat spacetime, similar to the vacuum case in \cite{asante2019holographic}, but rescaled in $\theta$. Thus, the mass dependence only becomes explicit when solving the equations of motion for the metric perturbations.

The equations of motion for $\lambda$ evaluated on background solutions yield the geodesic length,
\begin{equation}
    \ell = \frac{1}{2} \int^{r_{out}}_{r_{in}} dr \ \gamma_{\perp \perp}.
\end{equation}
This expression describes two possible scenarios. In the first, both an outer and an inner boundary exist, with geodesics spanning from a point on the outer boundary 
 $(r_{out},y)$ to the inner boundary point $(r_{in}, y)$.  In the second scenario, only an outer boundary is present, and geodesics extend from a boundary point to a point in the bulk at the center ($r=0,P_{r\rightarrow0}(y)$) where $P_{r\rightarrow0}(y)$ is a projection of the $y$ boundary point to a set of points described by $r=0$.

In order to solve for the metric perturbations and the Lagrange multiplier, we consider the equations of motion by varying the action (\ref{eqn:lagrangrmul}) with respect to the metric components,
\begin{equation}
\label{eqnmotion}
    \hat{G}^{ab} := V^{abcd} \gamma_{cd} + \frac{1}{2} G^{abcdef} \nabla_c \nabla_d \gamma_{ef} = \frac{1}{4} \frac{\lambda (y)}{\sqrt{h}} \delta^a_\perp \delta^b_\perp.
\end{equation}
Since we used Gaussian coordinates, we can write $\sqrt{g} = \sqrt{h}$. The contracted Bianchi identities, as calculated in \cite{asante2019holographic} for \eqref{eqnmotion}, guarantee three redundancies for the vacuum Einstein equations. This will allow us to solve for the three metric components $\gamma_{\perp \perp}$ and $\gamma_{\perp A}$ in terms of $\gamma_{AB}$ and $\lambda$. In the next section, we will derive the effective action for a surface given by the boundary of the solid torus.

\subsection{Solving for the metric perturbations} \label{twisted flat}


To solve the equations of motion \eqref{eqnmotion}, we Fourier transform the metric perturbations, 
\begin{equation}
     \gamma_{ab} (r, t, \theta) = \frac{1}{\sqrt{2 \pi \beta \mu}} \sum_{k'_t, k'_\theta} \gamma_{ab} (r,k_t,k_\theta) e^{i (\frac{\theta}{\mu}) k'_\theta}e^{i t k_t},
 \end{equation}
 with
   \begin{equation}
     \gamma_{ab} (r, k'_t, k'_\theta) = \frac{1}{\sqrt{2 \pi \beta \mu}} \int^{\beta/2}_{-\beta/2} dt \int^{\pi \mu}_{- \pi \mu} d\theta \; \gamma_{ab} (r,t,\theta) e^{-i (\frac{\theta}{\mu}) k'_\theta}e^{-i t k_t}.
 \end{equation}
 Here,
 the range for $\theta$ is $ -\pi \mu \leq \theta \leq \pi \mu $, which makes the conical deficit explicit.
We use the abbreviations $k_t = \frac{2 \pi}{\beta}(k'_t - \frac{\gamma}{2 \pi}k'_{\theta})$, and $k'_{\theta} , k'_t \in \mathbb{Z}$.

Taking the Fourier transform of the equations of motion \eqref{eqnmotion}, we solve for the lapse and shift components, $\gamma_{\perp \perp}$ and $\gamma_{\perp A}$, of the metric perturbations. This results in
\begin{flalign}
\label{eqn:lapse1}
    \gamma_{\perp \perp} = & 2 \partial_\perp \left( \frac{1}{2r} \left( \gamma_{\theta \theta} + \frac{k'^2_\theta}{\mu^2 k^2_t} \gamma_{tt} - 2\frac{k'_\theta}{\mu k_t} \gamma_{\theta t} \right)\right) && \\\nonumber
    = & 2\partial_\perp \xi^\perp,
\end{flalign}
\begin{flalign}
\label{eqn:lapse2}
    \gamma_{\perp \theta} = & i k'_\theta  \frac{1}{2r \mu} \left( \gamma_{\theta \theta} + \frac{k'^2_\theta}{\mu^2 k^2_t} \gamma_{tt} - 2\frac{k'_\theta}{\mu k_t} \gamma_{\theta t} \right) + r^2 \partial_\perp \left( \frac{i}{r^2} \left( \frac{k'_\theta}{2 \mu k^2_t} \gamma_{tt} - \frac{1}{k_t} \gamma_{\theta t} \right) \right) - ik'_\theta \lambda \frac{1}{4 \mu k_t^2} && \\\nonumber
     & =i \frac{k'_\theta}{\mu} \xi^\perp + r^2 \partial_\perp \xi^\theta - ik'_\theta \lambda \frac{1}{4 \mu k^2_t},
\end{flalign}
\begin{flalign}
\label{eqn:lapse3}
    \gamma_{\perp t} = & i k_t  \frac{1}{2r} \left( \gamma_{\theta \theta} + \frac{k'^2_\theta}{\mu^2 k^2_t} \gamma_{tt} - 2\frac{k'_\theta}{\mu k_t} \gamma_{\theta t} \right) + \partial_\perp \left( - \frac{i}{2k_t} \gamma_{tt} \right) - ik_t \lambda \frac{1}{4k_t^2} && \\\nonumber
     & =i k_t \xi^\perp +  \partial_\perp \xi^t - ik_t \lambda \frac{1}{4k^2_t}.
\end{flalign}
These expressions reduce to \eqref{eqn:diffeometric0} when $\lambda = 0$. Moreover, they match the vacuum solution reported in \cite{asante2019holographic}, with $k'_\theta$ replaced by $k'_\theta / \mu$, as expected, since the only difference between the geometries is the modified angular periodicity introduced by the conical defect. As we will see, this modification introduces an additional broken diffeomorphism, ultimately leading to the expected result for the massive BMS$_3$ character.

The solution for the diffeomorphism generating vector field $(\xi^\perp , \xi^A)$ is
\begin{equation}
\label{eqn:xisol}
    \xi^\perp = \frac{1}{2r} \left( \gamma_{\theta \theta} + \frac{k'^2_\theta}{\mu^2 k^2_t} \gamma_{tt} - 2\frac{k'_\theta}{\mu k_t} \gamma_{\theta t}  \right) ,
\end{equation}
\begin{equation}
    \xi^\theta = \frac{i}{r^2} \left( \frac{k'_\theta}{2 \mu k^2_t} \gamma_{tt} - \frac{1}{k_t} \gamma_{\theta t} \right),
\end{equation}
\begin{equation}
    \xi^t = - \frac{i}{2k_t} \gamma_{tt}.
\end{equation}
Taking the variation of the Lagrange multiplier of action \eqref{eqn:lagrangrmul}, we obtain 
\begin{equation}
    \rho = \frac{1}{2} \int^{r_2}_{r_1} dr \gamma_{\perp \perp} = \hat{\xi}^\perp (r_2) - \hat{\xi}^\perp (r_1),
\end{equation}
where
\begin{equation}
\label{xihat}
    \hat{\xi}^\perp = \xi^\perp -  \frac{1}{2 \Delta} \frac{\lambda}{\sqrt{h}} = \xi^\perp - \frac{1}{4k_t^2} \lambda.
\end{equation}
We note that in the case of a solid torus, \eqref{eqn:xisol} diverges at $r=0$. Furthermore, as seen in \eqref{xihat}, the $\lambda$ term is independent of $r$, which implies that for nonzero radius, the scalar field $\rho$ also becomes independent of $\lambda$. Consequently, $\lambda$ remains a free parameter, leaving no additional constraint to determine $\rho$. As argued in \cite{asante2019holographic}, this prevents the formulation of a well-defined dual boundary field theory for gravity since the $\lambda$-dependent terms cancel between the bulk and boundary contributions.

This issue is resolved by imposing smoothness conditions at the central axis of the twisted thermal flat spacetime. In the vacuum case, choosing the center of the torus is a gauge choice. In the massive case, however, this only holds partially. While there is still a symmetry in the time direction due to reparametrizations of the worldline, the radial position of the center is now fixed by the presence of the point particle. As we will see, this has direct implications for the smoothness conditions at $r=0$.

\subsection{Implementing smoothness conditions for the metric at $r = 0$} \label{sectionr0}

To impose smoothness conditions, we Taylor expand the metric perturbations around $r=0$ following \cite{asante2019holographic},
\begin{equation}
\begin{aligned}
\label{taylormetric}
    \gamma_{ab} = \ & a^{(0)}_{ab} + a^{(1)}_{ab} r + a^{(2)}_{ab} r^2 + \mathcal{O}(r^3) \hspace{3mm} \mathrm{for} \ ab = \perp \perp, \ tt, \ \perp t ; \\
    \gamma_{ab} = \ & a^{(1)}_{ab} r + a^{(2)}_{ab} r^2 + \mathcal{O}(r^3) \hspace{3mm}  \mathrm{for} \ ab = \perp \theta , \ \theta t ; \\
    \gamma_{\theta \theta} = \ & a^{(2)}_{\theta \theta} r^2 + \mathcal{O}(r^3).    
\end{aligned}
\end{equation}
To ensure finiteness at $r=0$ the following conditions need to be met: $a^{(n)}_{ab} = 0$ for $n < 0$, $a^{(0)}_{a \theta} = 0$ and $a^{(0)}_{\theta \theta}, a^{(1)}_{\theta \theta} = 0$. By matching these expressions with the Taylor expansion of \eqref{eqn:lapse1}, \eqref{eqn:lapse2} and \eqref{eqn:lapse3}, we obtain
\begin{equation}
\label{kthetaeqn0}
    \frac{k'^2_\theta}{\mu^2 k^2_t} a^{(0)}_{tt} = 0 ,
\end{equation}
\begin{equation}
\label{eqn:kcondition1}
    \left( 1 - \frac{\mu^2}{k'^2_\theta} \right) \left( \frac{k'^2_\theta}{\mu^2 k^2_t} a^{(1)}_{tt} - 2 \frac{k'_\theta}{\mu k_t} a_{\theta t}^{(1)} \right) = \frac{\lambda}{2k_t^2}.
\end{equation}
Equation \eqref{kthetaeqn0} ensures that $a^{(-1)}_{r \theta}$ and $a^{(-1)}_{r t}$ vanish while imposing the additional condition that $a^{(0)}_{tt} = 0$ for $k'_\theta \neq 0$.
On the other hand, \eqref{eqn:kcondition1} is well-defined for all $k'_\theta \neq 0$,
in contrast to \cite{asante2019holographic}, where $k'_\theta = \pm 1$ are problematic and they represent gauge modes. This demonstrates that $k'_\theta = \pm 1$ are no longer gauge modes but instead correspond to physical modes, representing broken diffeomorphisms caused by the presence of the point particle. This result aligns with the findings from the discrete setting, as discussed in Section \ref{gaugemodes}.

Imposing the condition $a^{(0)}_{tt} = 0$, we find that at $r=0$,
\begin{equation}
\begin{aligned}
    \xi^\perp (0) = &  \lim_{r\to 0} \frac{1}{2r} \left( \gamma_{\theta \theta} + \frac{k'^2_\theta}{k^2_t} \gamma_{tt} - 2\frac{k'_\theta}{k_t} \gamma_{\theta t} \right) \\ = & \frac{1}{2} \left( \frac{k'^2_\theta}{k^2_t} a^{(1)}_{tt} - 2 \frac{k'_\theta}{k_t} a_{\theta t}^{(1)} \right) \\ = &\frac{1}{4} \frac{k'^2_\theta}{(k'^2_\theta - \mu^2)} \frac{\lambda}{k_t^2}.
\end{aligned}
\end{equation}
We use this to solve the equation of motion for the Lagrange multiplier $\lambda$,
\begin{equation}
\label{eqn;rhowithzero}
    \rho = \frac{1}{2} \int^{r_{out}}_{0} dr \; \gamma_{\perp \perp} = \xi^\perp (r_{out}) - \xi^\perp (0) ,
\end{equation}
where
\begin{equation}
    \xi^\perp (r_{out}) = \frac{1}{2r_{out}} \left( \gamma_{\theta \theta} (r_{out}) + \frac{k'^2_\theta}{\mu^2 k^2_t} \gamma_{tt} (r_{out}) - 2\frac{k'_\theta}{\mu k_t} \gamma_{\theta t} (r_{out}) \right).
\end{equation}
Using these expressions, we find a solution for the Lagrange multiplier
\begin{equation}
\label{eqn:lambda}
    \lambda = 4k^2_t \left( 1 - \frac{\mu^2}{k'^2_\theta} \right) \left( \xi^\perp(r_{out}) - \rho \right).
\end{equation}
In order to deduce the dual boundary action, we evaluate the action \eqref{eqn:lagrangrmul} on this solution.

\subsection{Dual boundary action} 
Evaluating  the bulk term of the action \eqref{eqn:lagrangrmul} on the solutions to the equations of motion, we find
\begin{equation}
    \begin{aligned}
        - 8 \pi G S^{(2)}_{bulk} = & \frac{1}{4} \int_{\mathcal{M}} d^3x \sqrt{g} \gamma_{ab} \hat{G}^{ab} \\
        = & \frac{1}{16} \int_{\mathcal{M}} d^2y dr \gamma_{\perp \perp} (r, y) \lambda(y) \\
        = & \frac{1}{8} \int_{\partial \mathcal{M}} d^2y \lambda (y) (\xi^\perp (r_{out},y) - \xi^\perp(0,y)).
    \end{aligned}
\end{equation}
The boundary part of \eqref{eqn:lagrangrmul} consists of two parts: the boundary solution without $\lambda$ which is essentially $S^{(2)}_{HJ}$, and the boundary contribution proportional to the Lagrange multiplier $\lambda$ \cite{asante2019holographic}. This reads,
\begin{equation}
    -8 \pi G S^{(2)}_{bdry}= -8 \pi G S^{(2)}_{HJ} - \frac{1}{8} \int_{\partial \mathcal{M}} d^2y \lambda (y) \xi^\perp (r_{out}, y).
\end{equation}
However, the second term vanishes on the solution to the equations of motion \eqref{eqn;rhowithzero}. Therefore, the action \eqref{eqn:lagrangrmul} on-shell reduces to
\begin{equation}
\begin{aligned}
    -8 \pi G S^{(2)}_\lambda|_{sol} 
    = &  -8 \pi G S^{(2)}_{HJ}(r_{out})  - \frac{1}{8} \int_{\partial \mathcal{M}} d^2y \lambda(y) (\xi^\perp (0,y)).
\end{aligned}
\end{equation}
We note that the terms proportional to $\xi^\perp (r_{out}, y)$ cancel out. Substituting the solutions for $\lambda$ and $\xi^\perp(0,y)$ to the above expression, we obtain
\begin{equation}
\begin{aligned}
    -8 \pi G S^{(2)}_\lambda|_{sol} 
    = &  -8 \pi G S^{(2)}_{HJ}(r_{out}) - \frac{1}{2} \int_{\partial \mathcal{M}} d^2y \left[ (\xi^\perp (r_{out}) - \rho) k_t^2 \left( 1 - \frac{\mu^2}{k'^2_\theta} \right) (\xi^\perp (r_{out}) - \rho) \right] \\
\end{aligned}
\end{equation}
Taking the inverse Fourier transform and using \eqref{eqn:lambda}, we obtain
\begin{equation}
\begin{aligned}\label{eq:s2_lambda}
    -8 \pi G S^{(2)}_\lambda|_{sol} = &
    -8 \pi G S^{(2)}_{HJ}(r_{out}) + \frac{1}{2} \int_{\partial \mathcal{M}} d^2y \left[ (\xi^\perp (r_{out}) - \rho) \partial_t^2 \left( 1 + \frac{1}{\partial_\theta^2} \right) (\xi^\perp (r_{out}) - \rho) \right] \\
    = & -8 \pi G S^{(2)}_{HJ}(r_{out}) + \frac{1}{2} \int_{\partial \mathcal{M}} d^2y \xi^\perp (r_{out}) \partial_t^2 \left( 1 + \frac{1}{\partial_\theta^2} \right) \xi^\perp(r_{out}) \\
    & +\frac{1}{2} \int_{\partial \mathcal{M}} d^2y \left[ \rho \partial_t^2 \left(1 + \frac{1}{\partial_\theta^2} \right) \rho - 2 \rho \partial_t^2 \left(1 + \frac{1}{\partial_\theta^2} \right) \xi^\perp (r_{out}) \right] .
\end{aligned}
\end{equation}
Even though the above expression is the same as in the massless case as studied in \cite{asante2019holographic}, the mass-dependence is implicit in the field $\rho$ and the operator $1 / \partial_\theta^2$. 
Using \eqref{eqn:SHJVM}, this expression reduces to
\begin{equation}
\label{slambdafi}
\begin{aligned}
    - 8 \pi G S_\lambda^{(2)}|_{solu} = & - \frac{1}{4} \int_{\partial \mathcal{M}} d^2y \sqrt{h} \left( \rho \Delta \left( 1 + \frac{1}{\partial^2_\theta} \right) \rho - 2 \rho \left( 1 + \frac{1}{\partial^2_\theta} \right) \delta \ (^2R) \right)\\
    & + \frac{1}{4} \int_{\partial \mathcal{M}} d^2y \sqrt{h} \left( \xi^\perp \Delta \frac{1}{\partial_\theta^2} \xi^\perp - \xi^A D_{AB} \xi^B \right),
\end{aligned}
\end{equation}
where $\xi^\perp = \Delta^{-1} \delta \ (^2R) = -2^{-1}r \partial_t^{-2} \delta \ (^2R)$ accounts for the equations of motion for the field $\rho$.

The action \eqref{slambdafi} defines a boundary field theory for the field $\rho$. In particular, its on-shell action reproduces $S^{(2)}_{HJ}$ for a twisted thermal flat space. The first term of \eqref{slambdafi},

\begin{equation}
\label{eqn:dualaction}
    8 \pi G S'_\rho := \frac{1}{4} \int_{\partial \mathcal{M}} d^2y \sqrt{h} \left( \rho \Delta \left( 1 + \frac{1}{\partial^2_\theta} \right) \rho - 2 \rho \left( 1 + \frac{1}{\partial^2_\theta} \right) \delta \ (^2R) \right),
\end{equation}
is a non-local Liouville-type action and takes the exact same form as the effective dual boundary field action found in the massless case \cite{asante2019holographic}. This aligns with results obtained by taking the flat limit of Liouville theory \cite{Barnich:2012rz}.

Moreover, \eqref{eqn:dualaction} satisfies the conditions outlined in Section \ref{geolength} for a well-defined dual boundary field theory. However, in contrast to the massless case, $\rho$ and the non-local operator $(1 + \partial^{-2})$ have a  $\mu$-dependence. Furthermore, the modes $k'_\theta = 0$ of the non-local operator in \eqref{eqn:dualaction} remain gauge modes, while for $|k'_\theta| \geq 1$, they correspond to physical degrees of freedom in the massive case. This is the key feature that will be reflected in the one-loop determinant and ultimately reproduce the massive  BMS$_3$ character.

\subsection{One-loop determinant} \label{oneloopcont}
 We proceed to compute the one-loop determinant of \eqref{eqn:dualaction}, which by construction corresponds to the one-loop determinant of the gravitational action \eqref{actionpointpart}. To start,  we take the second derivative of \eqref{eqn:dualaction} with respect to $\rho$ and consider its Fourier transform
to obtain the Hessian,

\begin{equation}
\begin{aligned}
\label{hessianrho}
    8 \pi G S'^{(2)}_\rho = &  \prod^{\infty}_{k'_\theta=1} \prod^{\infty}_{k_t=0}\;  k^2_t \left( 1 - \frac{\mu^2}{k'^2_\theta} \right).
\end{aligned}
\end{equation}
Note that we have already taken $k'_\theta \geq 1$ to account for gauge modes. Following \cite{asante2019holographic}, we use a lattice regularization of the Laplace operator to compute \eqref{hessianrho}.
As extensively discussed in section \ref{section: Discrete}, we have a well-defined Regge discretization for the twisted thermal flat space. Therefore, we discretize the Laplacian using \eqref{omegafunction} to re-express $k_t^2$ and $k'^2_\theta$ in the discrete as follows,
\begin{equation}
    k'^2_\theta \rightarrow \Delta_\theta =   2 - 2\cos\left( \frac{2 \pi}{N_\theta} \kappa_\theta \right) ,
\end{equation}

\begin{equation}
    k_t^2 \rightarrow \Delta_t =  2 - 2\cos \left( \frac{2\pi}{N_T}(\kappa_t - \frac{\gamma}{2 \pi}\kappa_\theta)\right) ,
\end{equation}
where $\kappa_\theta = 0,...., N_A - 1$ and $\kappa_t = 0,...,N_T- 1$ with the same $N_A$ and $N_T$ used in the discrete case, and taking the limit  $N_T,N_A\rightarrow\infty$ removes the regularization. Normalizing these, we obtain

\begin{equation}
    k'^2_\theta \rightarrow \left( 2 - 2\cos\left( \frac{2 \pi}{N_A} \right) \right)^{-1} \left( 2 - 2\cos\left( \frac{2 \pi}{N_A} \kappa_\theta \right) \right) ,
\end{equation}

\begin{equation}
    k_t^2 \rightarrow \frac{N_T^2}{\beta^2} \left( 2 - 2\cos \left( \frac{2\pi}{N_T}(\kappa_t - \frac{\gamma}{2 \pi}\kappa_\theta) \right) \right).
\end{equation}
Substituting in \eqref{hessianrho} and computing the product over the $\kappa_\theta$ modes of the term in brackets, we obtain

\begin{equation}
\begin{aligned}
\label{onedetktheta}
    \prod^{N_A - 1}_{\kappa_\theta=1} \left[ 1 - 2 \left( \frac{ \mu^2 - \mu^2 \cos\left( \frac{2 \pi}{N_A} \right) }{ 2 - 2\cos\left( \frac{2 \pi}{N_A} \kappa_\theta \right)} \right) \right]
     = \prod^{N_A - 1}_{\kappa_\theta=1} \left[ 1 - \frac{2 X}{\Delta_\theta} \right] = f(X,N_\theta).
\end{aligned}
\end{equation}
The expression $f(X,N_\theta)$ is the same as in the discrete case \eqref{pochhammer} with $X = \mu^2 - \mu^2 \cos\left( \frac{2 \pi}{N_\theta} \right)$. We further compute the product over $\kappa_t$ and $\kappa_\theta$ modes for $k_t^2$,

\begin{equation}
\label{onedetkt}
    \prod^{N_A - 1}_{\kappa_\theta=1} \prod^{N_T - 1}_{\kappa_t=0} \frac{N_T^2}{\beta^2} \left( 2 - 2\cos \left( \frac{2\pi}{N_T}(\kappa_t - \frac{\gamma}{2 \pi}\kappa_\theta)\right) \right) = \frac{N_T^2}{\beta^2} \prod^{N_A - 1}_{\kappa_\theta=1} 2 - 2\cos(\gamma \kappa_\theta).
\end{equation}
As the final step in computing the full one-loop partition function, we evaluate the on-shell zeroth-order contribution to the gravitational action \eqref{actionpointpart}, which is given by
\begin{equation}
\begin{aligned}
\label{zeroorderS}
   S^{(0)}_p = & -\frac{ \beta}{4 G}\mu.
\end{aligned}
\end{equation}
Here, we have used the fact that the solution to the equations of motion is $R=-16 G \pi M \delta(\vec{r})$. This result is consistent with the one obtained in the discrete setting \eqref{szero}.

Putting together the zeroth-order contribution \eqref{zeroorderS} and the one-loop determinant of the Hessian given by \eqref{onedetktheta} and \eqref{onedetkt}, the resulting expression for the one-loop partition function is 
\begin{equation}
    \boxed{Z_{1-loop} \sim 
    \exp\left( \frac{1}{\hbar} \frac{\beta}{4G} \mu  \right) \prod^{(N_\theta-1)/2}_{\kappa_\theta = 1} \frac{1}{|1 - q^{\kappa_\theta}|^2},}
\end{equation}
where $q = \exp(i \gamma)$, and we have omitted lattice discretization constants and the boundary partition function, leaving only the essential contributions to the BMS$_3$ character.

This result matches the one-loop partition function obtained from Regge gravity \eqref{oneloopcontribution}. More importantly, in the limit $N_T,N_A\rightarrow\infty$, the one-loop determinant correctly reproduces the massive BMS$_3$ character. Similarly to the Regge gravity case, we have to take care of removing the lattice discretization since the product $|1 - q^{\kappa_\theta}|^{-2}$ is singular for $\gamma \times k$ an integer multiple of $2\pi$. Additionally, we recover the mass and central charge identification given by \eqref{masscharge}, further confirming that $S'_\rho$ is a dual boundary field action for 3D quantum gravity in a twisted thermal flat background with a conical defect.

\section{Discrete vs. continuum}\label{section: comp_w_discrete}
Besides the consistency checks we have made during the last section, we perform a consistency check by comparing the dual boundary action obtained directly in the continuum \eqref{eqn:dualaction} with the continuum limit of the action derived in the discrete case \eqref{hessboundmetric} building on the results of Sections \ref{section: Discrete} and \ref{section: Continuum}. To facilitate this comparison, we rewrite \eqref{eqn:dualaction} here 
\begin{equation}
\label{eqn:dualaction}
    8 \pi G S'_\rho = \frac{1}{4} \int_{\partial \mathcal{M}} d^2y \sqrt{h} \left( \rho \Delta \left( 1 + \frac{1}{\partial^2_\theta} \right) \rho - 2 \rho \left( 1 + \frac{1}{\partial^2_\theta} \right) \delta \ (^2R) \right).
\end{equation}
This effective boundary action has the same form as the one in the vacuum case reported in \cite{asante2019holographic}, the only difference being that if we express it in Fourier space,  $k'_\theta \geq 1$ instead of the $k'_\theta \geq 2$ result without a point particle. This is consistent with the fact that, in the vacuum case, diffeomorphism invariance in the radial direction is restored and the $k'_\theta \pm 1$ modes become gauge modes. Therefore, we can expect the resulting effective boundary action in the Regge gravity case to preserve the same form as in the vacuum case \cite{Bianca:2015}.

In Section \ref{section: Discrete}, we employed 3D Regge gravity as a tool to study the gravitational one-loop partition function by explicitly integrating over geometric fluctuations. However, translating physical quantities between discrete and continuum formulations in quantum gravity presents a significant challenge. To facilitate a direct comparison with continuum computations, we now analyze the continuum limit of various key quantities.
We begin by considering
\begin{equation}
            A = \varepsilon A_0 \mu \hspace{3mm},\hspace{3mm} T=\varepsilon T_0,
\end{equation}
where $A_0$ and $T_0$ are fixed and $\mu$ appears in the expression for $A$ to account for the conical deficit induced by the point particle.
Using geometrical relations and definitions given in Section \ref{section: Discrete}, we can derive the $\varepsilon$ dependence of the following quantities
\begin{equation}
    \begin{aligned}
        x & = \varepsilon^2\frac{A_0^2}{2R^2}\mu\\
        N_T & = \varepsilon^{-1}\frac{\beta}{T_0}\\
        N_A & = \varepsilon^{-1}\frac{2\pi R}{ A_0}+\mathcal{O}(\varepsilon^0)\\
        \omega_v & = 1+\varepsilon i\hat{v}-\varepsilon^2\frac{\hat{v}^2}{2}+\mathcal{O}(\varepsilon^3)\\
        \Delta_v & = \varepsilon^2\hat{v}^2
        +\mathcal{O}(\varepsilon^3)\\
        \omega_k & = 1+\varepsilon i \hat{k}-\varepsilon^2\frac{\hat{k}^2}{2}+\mathcal{O}(\varepsilon^3)\\
        \Delta_k & = \varepsilon^2\hat{k}^2
        +\mathcal{O}(\varepsilon^3) \\
         6V_\sigma & = \varepsilon^2A_0T_0R\; \mu+\mathcal{O}(\varepsilon^3),\label{continuumvariables}
    \end{aligned}
 \end{equation}   
 with $\hat{v}=\frac{2\pi T_0}{\beta}v$ and $\hat{k}=\frac{A_0}{R}k$. Thus, the continuum limit is achieved as taking $\varepsilon\rightarrow 0$. Geometrically, this corresponds to the limit when the size of the tetrahedra decreases while their number increases, leaving the volume of the twisted thermal torus fixed.\par
The boundary hessian \eqref{hessboundmetric} takes the form 
\begin{equation}
\begin{split}
    \Tilde{M}^h_\mathrm{b}(k,\nu)=-\frac{\varepsilon^{-2}R}{4A_0^3T_0\mu^2}\left(
\scalebox{0.8}{$\begin{array}{ccc}
 \frac{\hat{k}^4}{\hat{v}^2} &  \hat{k}^2 &  -2\frac{\hat{k}^3}{\hat{v}} \\
 \dots & \hat{v}^2 & -2\hat{k}\hat{v}\\
 \dots & \dots & 4\hat{k}^2\\ 
\end{array}$}
\right)+\frac{\varepsilon^{-2}}{4A_0T_0 \mu}\left(
\scalebox{0.8}{$\begin{array}{ccc}
 \frac{\hat{k}^2}{\hat{v}^2} & 0 &  -2\frac{\hat{k}}{\hat{v}} \\
 \dots & 0 & 0\\
 \dots & \dots & 4\\ 
\end{array}$}
\right)\\
+\varepsilon^{-2}\frac{A_0\; \mu}{16RT_0^3}\left(
\scalebox{0.8}{$\begin{array}{ccc}
 1 &  0 &  0 \\
 \dots & 0 & 0 \\
 \dots & \dots & 0\\ 
\end{array}$}
\right)+\mathcal{O}(\varepsilon^{-1}).
\end{split}
\end{equation}
From the continuum geometry computations, we know that the boundary field is equivalent to radial fluctuations of the diffeomorphism-generating vector field. Therefore, translating this into the discrete geometry setting, we focus on analyzing the radial displacements of vertices at the boundary. For a vertex $(s,n)$, such a displacement affects the length of the edges $(s,n)$ to $(s-1,n)$ in the angular direction and $(s,n)$ to $(s-1,n-1)$ in the diagonal direction. The deformation vector is given by 
\begin{equation}
    (n_{b,r})^t(k,\nu)=\frac{A}{\sqrt{6V_\sigma}}\hspace{1mm} \mathrm{sin}\left(\frac{\pi}{N_A}\mu\right)(0,1+\omega_k,1+\omega_v\omega_k)X_r(k,\nu),
\end{equation}
where $X_r$ is the radial component of a diffeomorphism-generating one-form (See Appendix B of \cite{Bianca:2015}). Using \eqref{eq:tometricvariables}, we can express this in terms of metric perturbations. This results in
\begin{equation}
    (n^h_{b,r})^t(k,\nu)=A\hspace{1mm} \mathrm{sin}\left(\frac{\pi}{N_A}\mu\right)(0,2(1+\omega_k),(1-\omega_v^{-1})\omega_v\omega_k)X_r(k,\nu).
\end{equation}
Finally, incorporating the scaling \eqref{continuumvariables}, this becomes
\begin{equation}
    (n^h_{b,r})^t(k,\nu)=\varepsilon^2 \frac{A_0^2}{R}\mu(0,2,0)X_r(k,\nu)+\mathcal{O}(\varepsilon^3).
\end{equation}
Therefore, the quadratic part of the effective boundary action in radial displacements in the continuum limit is given by
\begin{equation}
\begin{aligned}\label{discreteXr}
    8\pi G S_p^{(2)}[n^h_{b,\perp}] = &-\frac{1}{2}\sum_{k\geq 1,\nu\geq 0}\varepsilon^2 \frac{A_0}{RT_0}\;\hat{v}^2X_r(k,\nu)X_r(-k,-\nu)+\mathcal{O}(\varepsilon^3)
\end{aligned}
\end{equation}
We conclude that, in the continuum limit, the only way in which the effective boundary action of 3D gravity with a point particle differs from the vacuum case is through the $k = \pm 1$ modes. However, the functional form of these actions remains the same. This is consistent with the continuum geometry result \eqref{eqn:dualaction}.

\section{Conclusions and outlook}\label{section: Discussion}

In this work, we derived the one-loop partition function for 3D quantum gravity in a finite-radius thermal twisted flat space with a conical defect at its center. Its one-loop determinant reproduced the massive BMS$_3$ character found in \cite{Oblak:2016Oblackthesis}. The key difference from the vacuum case, where no conical defect is present, is that the $k=\pm 1$ modes of the gravitational action’s Hessian transition from gauge modes to physical ones both in the discrete and continuum geometry settings. As gauge modes correspond to null eigenvectors of the bulk Hessian associated with diffeomorphism invariance, this result clarifies that the extra mode in the massive BMS$_3$ character can be understood as a broken diffeomorphism in the radial direction due to the presence of a spinless particle.

Our main contribution is to provide a concrete and computable example where the breaking of diffeomorphism invariance in a discrete QG approach: Regge gravity, can be explicitly tracked. This allowed us to demonstrate how the emergence of physical degrees of freedom, such as the additional modes in the massive BMS$_3$ character, can be geometrically understood as a consequence of symmetry breaking. By reinterpreting smoothness conditions and gauge choices in light of this symmetry breaking, our results offer a clear benchmark for studying diffeomorphism breaking in discrete quantum gravity models and for testing the robustness of continuum dual descriptions.

In the discrete case, we obtained the Regge action for an inertial massive point particle in a 3D background and integrated out the bulk degrees of freedom, deriving an effective boundary action. The continuum limit was studied by taking the length of the temporal and angular edges to zero and the number of building blocks to infinity while keeping a finite-fixed-distance boundary. Additionally, as a consistency check, we expanded the massive case around the massless limit and confirmed that the result smoothly reduces to the known massless case \cite{Bianca:2015} in the limit $M\rightarrow0$.

In the continuum case, we constructed a dual boundary field theory following the path of \cite{asante2019holographic}, where the boundary field is precisely given by the length of geodesics starting at the worldline of the point particle and ending at the finite-distance boundary. The dual field theory turned out to be non-local and has exactly the same form as in the vacuum case. This demonstrates that the effect of a test point particle does not introduce new local physical degrees of freedom but instead modifies the global properties of spacetime. This is consistent with the results obtained in the discrete formulation and highlights the topological nature of the gravitational theory in both cases. Moreover, we confirmed the hypothesis stated in \cite{asante2019holographic} that introducing a point particle would break radial diffeomorphisms and modify the smoothness conditions imposed at $r=0$.

In both the discrete and continuum formulations of 3D gravity, studying the fluctuations of geodesic lengths anchored to the particle's worldline was essential for distinguishing between gauge and physical modes, ultimately leading to the recovery of the massive BMS$_3$ character. This demonstrates that discrete quantum gravity approaches can faithfully capture the dressing of classical point particles with gravitational degrees of freedom, as encoded in diffeomorphism-invariant metric fluctuations, providing a consistent interpretation of BMS$_3$ particles. In three-dimensional gravity, where there are no propagating gravitons, this dressing arises purely from bulk length fluctuations. In four dimensions, one would expect additional contributions from gravitons. Furthermore, since our results hold for finite-distance boundaries, this aligns with the edge mode interpretation proposed in \cite{Freidel:2023bnj}. Investigating whether our findings contribute to a formulation of edge modes from a path integral perspective remains an intriguing direction for future work.

A natural extension of this work is to generalize the discrete geometry computations to backgrounds with nonzero cosmological constant, particularly de Sitter (dS) and AdS spaces. Various discrete models in 3D homogeneously curved spacetimes provide tools for such an extension, such as deformed spinor networks \cite{Dupuis:2014qdeformed,Dupuis:2019yds}, the Turaev-Viro state sum \cite{Archer:1991tvspinfoam}, and discrete hypersurface deformation algebras \cite{Bonzom:2013tna}. These results could be compared with one-loop computations with matter in dS \cite{Castro:2023dxp} and AdS \cite{Benjamin:2020mfz}. Exploring connections with factorization and quantum deformation studies similarly to \cite{Mertens:2022ujr} would also be valuable in this context.  Additionally, for $\Lambda< 0$, there exist black hole solutions \cite{Banados:1992wn}, and a discrete geometry analysis of black hole thermodynamics, similar to \cite{Geiller:2013BTZspinfoams}, could help clarify the microscopic structure of black hole entropy from a discrete/simplicial geometry perspective. 

Another natural next step is to further analyze the dual boundary field theory associated with our effective boundary action. In AdS$_3$, dual theories for bulk conical defects have been studied \cite{Krasnov:2000PointParticleinADS}, where the insertion of conical defects in the bulk correspond to the insertion of vertex operators in the dual theory in the semiclassical limit. It would be interesting to check if this holds for our case. It has also been suggested that conical defects in the bulk can produce conformal anomalies in the boundary theory \cite{Arefeva:2016deficitnonconformal}. Understanding whether a similar phenomenon occurs in flat spacetimes would clarify the connection between bulk defects and boundary symmetries at finite distances.

From a group-theoretical perspective, an interesting extension of this work would be to construct the geometric action for flat space with a point particle, as in \cite{Barnich:2017jgw}, and compare it to the boundary action derived here. Additionally, \cite{Cotler:2024cia} demonstrated that the inclusion of matter in 3D gravity leads to the appearance of soft gravitons at null infinity. It would be interesting to compare this with our results to further investigate the role of gravitational dressings in the finite-distance regime.

Finally, since 3D Regge gravity is known to be the semiclassical limit of the Ponzano-Regge spinfoam model, an important direction for future research would be to compare our results with the ones obtained in the full quantum model \cite{Goeller:2019zpz}. Additionally, generalizing our study to four dimensions is a crucial step, especially given that the vacuum case in 4D has already been investigated in \cite{Asante:2018Vacuum4D,Asante:2021blx}.

\section*{Acknowledgements}
We thank Bianca Dittrich and Seth Asante for useful discussions and suggestions. AC was supported in part by the ANR-20-CE48-0018 “3DMaps” grant, the START-UP 2018 program (project number 740.018.017) funded by the Dutch Research Council (NWO), and Perimeter Institute for Theoretical Physics. Research at Perimeter Institute is supported by the Government of Canada through the Department of Innovation, Science, and Economic Development Canada, as well as by the Province of Ontario through the Ministry of Research, Innovation, and Science. We would also like to thank an anonymous reviewer who helped us to clarify to the reader the main contribution of this paper.


\section*{Appendix A: Second-order action with Lagrange multiplier}

The second-order action with Lagrange multiplier term is given by,
\begin{equation}
\begin{aligned}
    -8 \pi G S^{(2)}_\lambda = & \frac{1}{4} \int_{\mathcal{M}} d^3x \sqrt{g} \gamma_{ab} (V^{abcd} \gamma_{cd} + \frac{1}{2} G^{abcdef} \nabla_c \nabla_d \gamma_{ef}) + \\
    & \frac{1}{4} \int_{\partial \mathcal{M}} d^2y \sqrt{h} \gamma_{ab} ((B_1)^{abcd} \gamma_{cd} + (B_2)^{abcde} \nabla_c \gamma_{de}) + \\
    & \frac{1}{4} \int_{(\partial \mathcal{M})_{out}} d^2y \lambda(y) (\rho(y) - \ell[\gamma_{\perp \perp}])
\end{aligned}
\end{equation}
where
\begingroup
\begin{equation}
\begin{aligned}
    V^{abcd} = & \frac{1}{2} \left[ \frac{1}{2} (R - 2\Lambda) (g^{ab}g^{cd} - 2g^{ac}g^{bd}) - R^{ab}g^{cd} - g^abR^{cd} + 2(g^{ac}R^{bd} + g^{bc}R^{ad}) \right]
\end{aligned}
\end{equation}
\begin{equation}
\begin{aligned}
    G^{abefcd} = & g^{ab}g^{ec}g^{fd} + g^{ac}g^{bd}g^{ef} + g^{ae}g^{bf}g^{cd} - g^{ab}g^{ef}g^{cd} - g^{af}g^{bd}g^{ec} - g^{ac}g^{bf}g^{ed}
\end{aligned}
\end{equation}
\begin{equation}
\begin{aligned}
    B_1^{abcd} = & \frac{1}{2} (Kh^{ab} - K^{ab})g^{cd} - h^{ac}h^{bd}K - h^{ab}K^{cd} + h^{ac}K^{bd} + h^{bc}K^{ad}
\end{aligned}
\end{equation}
\begin{equation}
\begin{aligned}
    B_2^{abecd} = & \frac{1}{2} ((h^{ae}h^{bd} - h^{ab}h^{ed})n^c + (h^{ac}h^{be} - h^{ab}h^{ce})n^d - (h^{ac}h^{bd} - h^{ab}h^{cd})n^e.
\end{aligned}
\end{equation}
\endgroup
and $n^a = (-1,0,0)$ is an outward-pointing normal vector associated to the outer boundary.

\bibliographystyle{ieeetr}
\bibliography{references}

\begin{thebibliography}{10}

\bibitem{Brown:1986nw}
J.~D. Brown and M.~Henneaux, ``{Central Charges in the Canonical Realization of
  Asymptotic Symmetries: An Example from Three-Dimensional Gravity},'' {\em
  Commun. Math. Phys.}, vol.~104, pp.~207--226, 1986.

\bibitem{Goddard:1986ee}
P.~Goddard, A.~Kent, and D.~I. Olive, ``{Unitary Representations of the
  Virasoro and Supervirasoro Algebras},'' {\em Commun. Math. Phys.}, vol.~103,
  pp.~105--119, 1986.

\bibitem{Barnich:2001jy}
G.~Barnich and F.~Brandt, ``{Covariant theory of asymptotic symmetries,
  conservation laws and central charges},'' {\em Nucl. Phys. B}, vol.~633,
  pp.~3--82, 2002.

\bibitem{Bondi:1962px}
H.~Bondi, M.~G.~J. van~der Burg, and A.~W.~K. Metzner, ``{Gravitational waves
  in general relativity. 7. Waves from axisymmetric isolated systems},'' {\em
  Proc. Roy. Soc. Lond. A}, vol.~269, pp.~21--52, 1962.

\bibitem{Sachs:1962BMSORIGINAL}
R.~K. Sachs, ``{Gravitational waves in general relativity. 8. Waves in
  asymptotically flat space-times},'' {\em Proc. Roy. Soc. Lond.}, vol.~A270,
  pp.~103--126, 1962.

\bibitem{Oblak:2016Oblackthesis}
B.~Oblak, {\em {BMS Particles in Three Dimensions}}.
\newblock PhD thesis, Brussels U., 2016.

\bibitem{Witten:19883DGRAVITY}
E.~Witten, ``{(2+1)-Dimensional Gravity as an Exactly Soluble System},'' {\em
  Nucl. Phys.}, vol.~B311, p.~46, 1988.

\bibitem{Carlip:1998BOOK}
S.~Carlip, {\em {Quantum gravity in 2+1 dimensions}}.
\newblock Cambridge Monographs on Mathematical Physics, Cambridge University
  Press, 2003.

\bibitem{Oeckl:2003vu}
R.~Oeckl, ``{A 'General boundary' formulation for quantum mechanics and quantum
  gravity},'' {\em Phys. Lett. B}, vol.~575, pp.~318--324, 2003.

\bibitem{Dittrich:2018xuk}
B.~Dittrich, C.~Goeller, E.~R. Livine, and A.~Riello, ``{Quasi-local
  holographic dualities in non-perturbative 3d quantum gravity},'' {\em Class.
  Quant. Grav.}, vol.~35, no.~13, p.~13LT01, 2018.

\bibitem{Goeller:2019zpz}
C.~Goeller, E.~R. Livine, and A.~Riello, ``{Non-Perturbative 3D Quantum
  Gravity: Quantum Boundary States and Exact Partition Function},'' {\em Gen.
  Rel. Grav.}, vol.~52, no.~3, p.~24, 2020.

\bibitem{Freidel:2021ajp}
L.~Freidel, C.~Goeller, and E.~R. Livine, ``{The quantum gravity disk: Discrete
  current algebra},'' {\em J. Math. Phys.}, vol.~62, no.~10, p.~102303, 2021.

\bibitem{Freidel:2023bnj}
L.~Freidel, M.~Geiller, and W.~Wieland, {\em {Corner Symmetry and Quantum
  Geometry}}, pp.~1--36.
\newblock 2024.

\bibitem{ambjorn2009geometry}
J.~Ambjorn, M.~Carfora, and A.~Marzuoli, {\em The Geometry of Dynamical
  Triangulations}.
\newblock Lecture Notes in Physics Monographs, Springer Berlin Heidelberg,
  2009.

\bibitem{Ambjorn:2000dja}
J.~Ambjorn, J.~Jurkiewicz, and R.~Loll, ``{Nonperturbative 3-D Lorentzian
  quantum gravity},'' {\em Phys. Rev. D}, vol.~64, p.~044011, 2001.

\bibitem{Barrett:2008wh}
J.~W. Barrett and I.~Naish-Guzman, ``{The Ponzano-Regge model},'' {\em Class.
  Quant. Grav.}, vol.~26, p.~155014, 2009.

\bibitem{Jafferis:2024jkb}
D.~L. Jafferis, L.~Rozenberg, and G.~Wong, ``{3d Gravity as a random
  ensemble},'' 7 2024.

\bibitem{Regge:1961px}
T.~Regge, ``{GENERAL RELATIVITY WITHOUT COORDINATES},'' {\em Nuovo Cim.},
  vol.~19, pp.~558--571, 1961.

\bibitem{dittrich2012path}
B.~Dittrich and S.~Steinhaus, ``{Path integral measure and triangulation
  independence in discrete gravity},'' {\em Phys. Rev. D}, vol.~85, p.~044032,
  2012.

\bibitem{Bianca:2015}
V.~Bonzom and B.~Dittrich, ``{3D holography: from discretum to continuum},''
  {\em JHEP}, vol.~03, p.~208, 2016.

\bibitem{asante2019holographic}
S.~K. Asante, B.~Dittrich, and F.~Hopfmueller, ``Holographic formulation of 3d
  metric gravity with finite boundaries,'' {\em Universe}, vol.~5, no.~8,
  p.~181, 2019.

\bibitem{Barnich:2015Partitionfunctionatoneloop}
G.~Barnich, H.~A. Gonzalez, A.~Maloney, and B.~Oblak, ``{One-loop partition
  function of three-dimensional flat gravity},'' {\em JHEP}, vol.~04, p.~178,
  2015.

\bibitem{Maloney:2007ud}
A.~Maloney and E.~Witten, ``{Quantum Gravity Partition Functions in Three
  Dimensions},'' {\em JHEP}, vol.~02, p.~029, 2010.

\bibitem{Giombi:2008vd}
S.~Giombi, A.~Maloney, and X.~Yin, ``{One-loop Partition Functions of 3D
  Gravity},'' {\em JHEP}, vol.~08, p.~007, 2008.

\bibitem{Banados:1992wn}
M.~Banados, C.~Teitelboim, and J.~Zanelli, ``{The Black hole in
  three-dimensional space-time},'' {\em Phys. Rev. Lett.}, vol.~69,
  pp.~1849--1851, 1992.

\bibitem{Krasnov:2000PointParticleinADS}
K.~Krasnov, ``{3-D gravity, point particles and Liouville theory},'' {\em
  Class. Quant. Grav.}, vol.~18, pp.~1291--1304, 2001.

\bibitem{Maxfield:2020ale}
H.~Maxfield and G.~J. Turiaci, ``{The path integral of 3D gravity near
  extremality; or, JT gravity with defects as a matrix integral},'' {\em JHEP},
  vol.~01, p.~118, 2021.

\bibitem{Benjamin:2020mfz}
N.~Benjamin, S.~Collier, and A.~Maloney, ``{Pure Gravity and Conical
  Defects},'' {\em JHEP}, vol.~09, p.~034, 2020.

\bibitem{Deser:19833Dpointparticlessolution}
S.~Deser, R.~Jackiw, and G.~'t~Hooft, ``{Three-Dimensional Einstein Gravity:
  Dynamics of Flat Space},'' {\em Annals Phys.}, vol.~152, p.~220, 1984.

\bibitem{hartle1981boundary}
J.~B. Hartle and R.~Sorkin, ``{Boundary Terms in the Action for the Regge
  Calculus},'' {\em Gen. Rel. Grav.}, vol.~13, pp.~541--549, 1981.

\bibitem{Rocek:1982tj}
M.~Rocek and R.~M. Williams, ``{The Quantization of Regge Calculus},'' {\em Z.
  Phys. C}, vol.~21, p.~371, 1984.

\bibitem{Dittrich:2007HessianFirst}
B.~Dittrich, L.~Freidel, and S.~Speziale, ``{Linearized dynamics from the
  4-simplex Regge action},'' {\em Phys. Rev.}, vol.~D76, p.~104020, 2007.

\bibitem{Bahr:2009Reggeactionlinearization}
B.~Bahr and B.~Dittrich, ``{Regge calculus from a new angle},'' {\em New J.
  Phys.}, vol.~12, p.~033010, 2010.

\bibitem{Oblak:2015BMS}
B.~Oblak, ``{Characters of the BMS Group in Three Dimensions},'' {\em Commun.
  Math. Phys.}, vol.~340, no.~1, pp.~413--432, 2015.

\bibitem{Barnich:2012rz}
G.~Barnich, A.~Gomberoff, and H.~A. Gonz\'alez, ``{Three-dimensional
  Bondi-Metzner-Sachs invariant two-dimensional field theories as the flat
  limit of Liouville theory},'' {\em Phys. Rev. D}, vol.~87, no.~12, p.~124032,
  2013.

\bibitem{Dupuis:2014qdeformed}
M.~Dupuis, F.~Girelli, and E.~R. Livine, ``{Deformed Spinor Networks for Loop
  Gravity: Towards Hyperbolic Twisted Geometries},'' {\em Gen. Rel. Grav.},
  vol.~46, no.~11, p.~1802, 2014.

\bibitem{Dupuis:2019yds}
M.~Dupuis, E.~R. Livine, and Q.~Pan, ``{$q$-deformed 3D Loop Gravity on the
  Torus},'' {\em Class. Quant. Grav.}, vol.~37, no.~2, p.~025017, 2020.

\bibitem{Archer:1991tvspinfoam}
F.~Archer and R.~M. Williams, ``{The Turaev-Viro state sum model and
  three-dimensional quantum gravity},'' {\em Phys. Lett. B}, vol.~273,
  pp.~438--444, 1991.

\bibitem{Bonzom:2013tna}
V.~Bonzom and B.~Dittrich, ``{Dirac\textquoteright{}s discrete hypersurface
  deformation algebras},'' {\em Class. Quant. Grav.}, vol.~30, p.~205013, 2013.

\bibitem{Castro:2023dxp}
A.~Castro, I.~Coman, J.~R. Fliss, and C.~Zukowski, ``{Keeping matter in the
  loop in dS$_{3}$ quantum gravity},'' {\em JHEP}, vol.~07, p.~120, 2023.
\newblock [Erratum: JHEP 09, 004 (2024)].

\bibitem{Mertens:2022ujr}
T.~G. Mertens, J.~Sim\'on, and G.~Wong, ``{A proposal for 3d quantum gravity
  and its bulk factorization},'' {\em JHEP}, vol.~06, p.~134, 2023.

\bibitem{Geiller:2013BTZspinfoams}
M.~Geiller and K.~Noui, ``{BTZ Black Hole Entropy and the Turaev-Viro model},''
  {\em Annales Henri Poincare}, vol.~16, no.~2, pp.~609--640, 2015.

\bibitem{Arefeva:2016deficitnonconformal}
I.~{\relax Ya}. Aref'eva and M.~A. Khramtsov, ``{AdS/CFT prescription for
  angle-deficit space and winding geodesics},'' {\em JHEP}, vol.~04, p.~121,
  2016.

\bibitem{Cotler:2024cia}
J.~Cotler, K.~Jensen, S.~Prohazka, M.~Riegler, and J.~Salzer, ``{Soft gravitons
  in three dimensions},'' 11 2024.

\bibitem{Asante:2018Vacuum4D}
S.~K. Asante, B.~Dittrich, and H.~M. Haggard, ``{Holographic description of
  boundary gravitons in (3+1) dimensions},'' {\em JHEP}, vol.~01, p.~144, 2019.

\bibitem{Asante:2021blx}
S.~K. Asante and B.~Dittrich, ``{Perfect discretizations as a gateway to
  one-loop partition functions for 4D gravity},'' {\em JHEP}, vol.~05, p.~172,
  2022.

\end{thebibliography}

\end{document}